\documentclass[a4paper,fleqn,usenatbib]{mnras}

\usepackage{amsmath}	
\usepackage{txfonts}
\usepackage{mathrsfs}
\usepackage[T1]{fontenc}
\usepackage{ae,aecompl}

\usepackage{graphicx}	
\usepackage{amssymb}	
\usepackage{multirow}
\usepackage{hyperref}
\hypersetup{
    colorlinks,
    citecolor=blue,
    filecolor=black,
    linkcolor=blue,
    urlcolor=black
}

\newcommand{\fesc}{\ifmmode{f_{\rm esc}}\else{$f_{\rm esc}$}\fi}
\newcommand{\fescs}{\ifmmode{f_{\rm esc}^\star}\else{$f_{\rm esc}^\star$}\fi}
\newcommand{\kms}{\ifmmode{{\;\rm km~s^{-1}}}\else{km~s$^{-1}$}\fi}
\newcommand{\fgas}{\ifmmode{{f_{\rm gas}}}\else{$f_{\rm gas}$}\fi}
\newcommand{\cubecm}{\ifmmode{{\rm cm^{-3}}}\else{cm$^{-3}$}\fi}
\newcommand{\ztwo}{\ifmmode{{\rm [Z_2/H]}}\else{[Z$_2$/H]}\fi}
\newcommand{\zthree}{\ifmmode{{\rm [Z_3/H]}}\else{[Z$_3$/H]}\fi}
\newcommand{\lsim}{\lower0.3em\hbox{$\,\buildrel <\over\sim\,$}}
\newcommand{\gsim}{\lower0.3em\hbox{$\,\buildrel >\over\sim\,$}}

\newcommand{\sfr}{\ifmmode{\textrm{M}_\odot \,\textrm{yr}^{-1} \,\textrm{Mpc}^{-3}}\else{M$_\odot$ yr$^{-1}$ Mpc$^{-3}$}\fi}
\newcommand{\hsfr}{\ifmmode{\textrm{M}_\odot\, \textrm{yr}^{-1}}\else{M$_\odot$ yr$^{-1}$}\fi}

\newcommand{\eavg}{\ifmmode{\langle E_\gamma \rangle}\else{$\langle E_\gamma \rangle$}\fi}

\newcommand{\enzo}{{\sc enzo}}
\newcommand{\yt}{{\sc yt}}
\newcommand{\moray}{{\sc enzo+moray}}
\newcommand{\Ms}{\ifmmode{M_\odot}\else{$M_\odot$}\fi}
\newcommand{\vrms}{\ifmmode{v_{\rm rms}}\else{$v_{\rm rms}$}\fi}

\newcommand{\tvir}{\ifmmode{T_{\rm{vir}}}\else{$T_{\rm{vir}}$}\fi}
\newcommand{\mvir}{\ifmmode{M_{\rm{vir}}}\else{$M_{\rm{vir}}$}\fi}
\newcommand{\rvir}{\ifmmode{r_{\rm{vir}}}\else{$r_{\rm{vir}}$}\fi}

\newcommand{\jj}{\ifmmode{J_{21}}\else{$J_{21}$}\fi}
\newcommand{\flw}{\ifmmode{F_{LW}}\else{$F_{LW}$}\fi}
\newcommand{\kph}{\ifmmode{k_{\rm ph}}\else{$k_{\rm ph}$}\fi}

\newcommand{\zsun}{\ifmmode{\rm\,Z_\odot}\else{$\rm\,Z_\odot$}\fi}

\newcommand{\hi}{H {\sc i}}
\newcommand{\hii}{H {\sc ii}}
\newcommand{\hei}{He {\sc i}}
\newcommand{\heii}{He {\sc ii}}
\newcommand{\heiii}{He {\sc iii}}
\newcommand{\nhi}{\ifmmode{N_{\rm HI}}\else{$N_{\rm HI}$}\fi}

\newcommand{\oiii}{[O {\sc iii}]}
\newcommand{\ciii}{C {\sc iii}]}
\newcommand{\sii}{S {\sc ii}}
\newcommand{\nii}{N {\sc ii}}

\def\eps@scaling{1.0}%
\newcommand\epsscale[1]{\gdef\eps@scaling{#1}}%
\newcommand\plotone[1]{%
 \centering 
 \leavevmode 
 \includegraphics[width={\eps@scaling\columnwidth}]{#1}%
}%
\newcommand\plottwo[2]{%
 \centering 
 \includegraphics[width={\eps@scaling\columnwidth}]{#1}%
 \hfil 
 \includegraphics[width={\eps@scaling\columnwidth}]{#2}%
}%

\title[Mock Spectra of First Galaxies]{First Light: exploring the Spectra of High-Redshift Galaxies in the Renaissance Simulations}

\author[K. S. S. Barrow et al.]{Kirk S. S. Barrow$^1$\thanks{e-mail:
    kssbarrow@gatech.edu}, John H. Wise$^1$, Michael L. Norman$^2$, Brian W. O'Shea$^3$, \newauthor and Hao Xu$^{2,4}$\\
  $^{1}$ Center for Relativistic Astrophysics, Georgia Institute of
  Technology, 837 State Street, Atlanta, GA
  30332, USA\\
  $^{2}$ CASS, University of California, San Diego, 9500 Gilman Drive, La Jolla, CA 92093, USA\\
  $^3$ 
Department of Computational Mathematics, Science and Engineering, Department of Physics and Astronomy, \\
  \ \ \ and National Superconducting Cyclotron Laboratory, Michigan State University, East Lansing, MI 48824, USA\\
  $^4$ IBM, New Orchard Road, Armonk, NY 10504, USA
}
\pagerange{4863--4878} \pubyear{2017} \volume{469(4)}

\date{Accepted May 15, 2017. Received May 1, 2017; in original form January 10, 2017}

\begin{document}
\label{firstpage}
\maketitle

\begin{abstract}
  
We present synthetic observations for the first generations of galaxies in the Universe and make predictions for future deep field observations for redshifts greater than 6. Due to the strong impact of nebular emission lines and the relatively compact scale of \hii{} regions, high resolution cosmological simulations and a robust suite of analysis tools are required to properly simulate spectra. We created a software pipeline consisting of {\sc FSPS}, {\sc Hyperion}, {\sc Cloudy} and our own tools to generate synthetic IR observations from a fully three-dimensional arrangement of gas, dust, and stars. Our prescription allows us to include emission lines for a complete chemical network and tackle the effect of dust extinction and scattering in the various lines of sight. We provide spectra, 2-D binned photon imagery for both HST and JWST IR filters, luminosity relationships, and emission line strengths for a large sample of high redshift galaxies in the Renaissance Simulations. Our resulting synthetic spectra show high variability between galactic halos with a strong dependence on stellar mass, metallicity, gas mass fraction, and formation history. haloes with the lowest stellar mass have the greatest variability in \oiii{}/H$\beta$, \oiii{} and \ciii{} while haloes with higher masses are seen to show consistency in their spectra and \oiii{} equivalent widths (EW) between 1\AA\ and 10\AA. Viewing angle accounted for three-fold difference in flux due to the presence of ionized gas channels in a halo. Furthermore, JWST colour plots show a discernible relationship between redshift, colour, and mean stellar age.

\end{abstract}

\begin{keywords} 
(cosmology:) dark ages,reionization,first stars--techniques: spectroscopy--photometric--methods: radiative transfer--numerical--observational
\end{keywords}

\section{Introduction and Background}

The frontier of observations and cosmological simulations has been mostly defined by the limitations of hardware and the fidelity of modelling methods. The epoch of reionization (EoR; $6 \la z \la 15$) is both the farthest back space-based observation has seen and the furthest forward large, high-fidelity radiative transfer ``first-galaxy'' simulations have reached, generating a thin region of overlap where real and synthetic observations may be compared and predicted in greater detail.

\subsection{EoR Galaxy Observations}

The bright UV portion of a young galactic spectrum in the rest frame at $z > 6$ corresponds to the optical and near infrared portion of a spectrum for a present day observer. Thus only telescopes calibrated to observe these and higher wavelengths are appropriate for high-redshift observations. The current cache of operating IR-capable space telescopes include the Wide-field Infrared Survey Explorer (WISE), the Spitzer Space Telescope, and the Hubble Space Telescope (HST). Of the three, the HST's deep field surveys using the UVIS/IR Wide Field Camera (WFC3) provide the most useful data for studies of this epoch. 

The Hubble Ultra Deep Field \citep[HUDF;][]{2006AJ....132.1729B} is a culmination of many years of observational campaigns.  Several groups have uncovered hundreds of galaxies with redshifts greater than 5 through the Lyman break ``drop out'' technique \citep{1996AJ....112..352S} that uses photometry with appropriate colour-colour selection rules to high-redshift candidates \citep{2006AJ....132.1729B}.  In its initial 2005 HUDF campaign, the Hubble Advanced Camera for Surveys (ACS) used four optical filters to identify galaxies with redshifts up to 7.5, showing that the galaxy number density further declines with redshift \citep[e.g.][]{2005ApJ...626..666M, 2007ApJ...670..928B, 2007ApJ...671.1212O}.

The HUDF was supplemented with WFC3 observations in 2009, confirming earlier results and extending its reach beyond $z \simeq 7.5$ \citep[e.g.,][]{2010MNRAS.403..960M}.  By combining results from optical and near-infrared imaging, this survey used SED fitting to catalogue and confirm 49 candidate objects between redshifts 6 and 9. A subsequent campaign in 2012 with longer exposure times and through more filters, dubbed the Hubble Extreme Deep Field \citep[XDF;][]{2013ApJS..209....3K}, produced candidates between redshifts 8.5 and 12 \citep{2013ApJ...763L...7E}. The luminosity functions determined from the XDF galaxies show an exponential decay with increasing redshift until approximately $z = 8$ \citep[e.g.][]{2013ApJS..209....3K, 2014ApJ...795..126B, 2016PASA...33...37F} followed by steeper drop off subject to great uncertainty due to the limited number of observations and difficulty confirming their redshifts. Specifically at both the brightest end ($\rm{M_{1600}} \leq -22$) and the faint end ($\rm{M_{1600}} \geq -17$), the uncertainties between and within fits grow to a several orders of magnitude differences in the number density of objects \citep{2015MNRAS.452.1817B}.

The current state of the art is the ongoing Hubble Frontier Fields survey \citep{2016arXiv160506567L} which combines ACS, WFC3 and catalogues of galaxy clusters \citep{1989ApJS...70....1A,2001ApJ...553..668E} to resolve gravitationally lensed objects in their vicinity on the sky as seen from the Earth. The survey is designed to probe six clusters to an observed $\rm{AB_{mag}} \sim 29$ which may be magnified 10--100 times in fortuitous alignments with foreground strongly-lensing galaxy clusters.

\subsection{EoR Galaxy Simulations}

Cosmological simulations apply numerical and empirical subgrid prescriptions to a set of cosmological initial conditions at high redshift to produce a physically plausible representation of the Universe. To that end several codes that include star formation and feedback, radiative transfer, and hydrodynamics routines have been developed. However, as the routines became more physically representative, their demands on computational infrastructure constrain their resolution and scope. Where small-scale and zoom-in simulations are capable of astounding spatial resolution, accurate accounting for physical phenomena in large-scale cosmological simulations implies the propagation of radiation and the interaction of processes that occur simultaneously on a variety of time and spatial scales from atomic to cosmological.

To achieve appropriate and efficient scale, cosmological simulations generally employ either smooth particle hydrodynamics (SPH) or adaptive mesh refinement (AMR). In addition to the body of work using SPH or AMR, some simulations employ moving mesh codes such as {\sc Arepo} \citep{2010MNRAS.401..791S} and meshless codes such as {\sc Gizmo} \citep{2015MNRAS.450...53H}. SPH simulation codes such as {\sc Gadget} \citep{2005MNRAS.364.1105S} and {\sc Gasoline} \citep{2004NewA....9..137W} initialize mass as a distribution of particles which evolve hydrodynamically into high and low density regions. Fluid dynamics and other calculations are then smoothed over a fixed number of nearby particles, generating a density-dependent spatial resolution for physical calculations. Conversely, AMR codes such as \enzo{} \citep{2014ApJS..211...19B} and {\sc Ramses} \citep{2002A&A...385..337T} initialize a uniform mesh of initial quantities and attributes where calculations of physical processes are calculated within and between each constant-quantity cell of the mesh. Based on one or more user-defined parameters which may include density, cells are further divided into smaller regions as needed to track regions that require higher spatial resolution. Studies indicate that the results of these two methods are similar with discrepancies owning mostly to their sub-grid physics models \citep{2005ApJS..160....1O,2014ApJS..210...14K,2016ApJ...833..202K}.

Examples of high-redshift galaxy simulations with high physical fidelity, radiative transfer, and high resolution are represented by the Renaissance Simulations \citep{2015ApJ...807L..12O,2016arXiv160407842X} used in this work, the Aurora Simulation \citep{2016arXiv160300034P}, the BlueTides Simulation \citep{2016MNRAS.455.2778F}, the First Billion Years Project \citep{2015MNRAS.451.2544P}, and a recent radiative transfer-modified {\sc Ramses} simulation by \citet{2016arXiv161201786K}.

\subsection{Synthetic Observations}


Using the HUDF and two different sets of parallel fields, \citet{2014ApJ...795..126B} and \citet{2015ApJ...810...71F} constrained luminosity functions up to redshift 8 and down to $\rm{M_{1600}} \simeq -17$ within the EoR.  By using gravitational lensing, the luminosity functions from the Frontier Fields extend the steep faint-end slope out to $M_{\rm UV} \simeq -12.5$ at $z \sim 6$ and $-14$ at $z \sim 7-8$ with no apparent flattening \citep{2015ApJ...814...69A, 2016arXiv160406799L, 2016arXiv161000283B}.  The luminosity function of objects in the Renaissance Simulations \citep{2015ApJ...807L..12O} likewise converged on $\rm{M_{1600}} \simeq -17$ from the faint end of the luminosity function, closing a long-standing gap between the dimmest observable EoR objects observed and the brightest objects produced in ``first galaxy'' cosmological simulations. This convergence is timely for the creation of synthetic observations that can be directly compared to images from the HUDF and the Frontier Fields.

Unfortunately, the characterization of observations of individual objects at high redshift is sometimes difficult due to the complex geometry of emission from early galaxies and their distance. For example, objects such as CR7 were initially characterized as a possible Population III galaxy \citep{2015ApJ...808..139S} or evidence of a direct collapse black hole \citep{2015MNRAS.453.2465P,2016MNRAS.460.3143S,2016ApJ...823...74D,2016MNRAS.460.4003A}, but further study produced evidence to dispute both claims \citep{2016arXiv160900727B}. Due to stellar feedback and a high merger rate, the morphology of large early galaxies is irregular and exhibit high variability with respect to the observer's viewing angle. These galaxies have bursty star formation and relatively high UV escape fractions \citep{2016arXiv160407842X}, making for a non-trivial star formation history. Observations from the Atacama Large Millimetre/Submillimetre Array (ALMA) further suggest that galaxies at high redshift ($6 \lsim z \lsim 8$) may exhibit higher metallicities and dust content that previously thought \citep{2015ApJ...807..180W,2016arXiv160706772A}, emphasizing the need for a robust dust model. 

To generate synthetic observations of objects like CR7, a full 3D model of stellar populations as well as gas and dust extinction is therefore desired. Furthermore, more capable space telescopes like the forthcoming James Webb Space Telescope (JWST) are required to better study these objects and explore a more statistically significant sample size of early-universe galaxies.

Telescopes observing this epoch are limited by the spectral and pixel resolution of their detectors and can thus produce photometry and low signal to noise spectra data that requires extensive fitting to templates produced by theory. On the other hand, simulations are particularly suited to exactly calculate values that can only be empirically determined from observations such as a star formation history, mass to light ratios, column densities, and high-resolution spectra. This work seeks to provide a suite of such measures for the large sample of galaxies in the Renaissance Simulations to provide context for observations as well as demonstrate the variability that comes with viewing angles and formation history. Similarly we hope to use observations to provide context and constrain uncertainties in the subgrid models used in cosmological simulations.

We provide an overview of the physics modelling and simulation setup of the Renaissance Simulations and our own post-processing radiative transfer routines through a dusty medium in Section \ref{sec:EnzoStar} and discuss our methods for the production of synthetic observations and measures. We then analyse two individual galaxies and then provide aggregate measures such as composite spectra and line ratios for the entire sample in Section \ref{AggHalo}. Finally we discuss our results and compare it to recent work on synthetic observations and emission lines in Section \ref{discussion}.

\section{Research Methods}

\label{sec:EnzoStar}

\subsection{Simulation Techniques}

\begin{figure}
\begin{center}
\includegraphics[scale=.30]{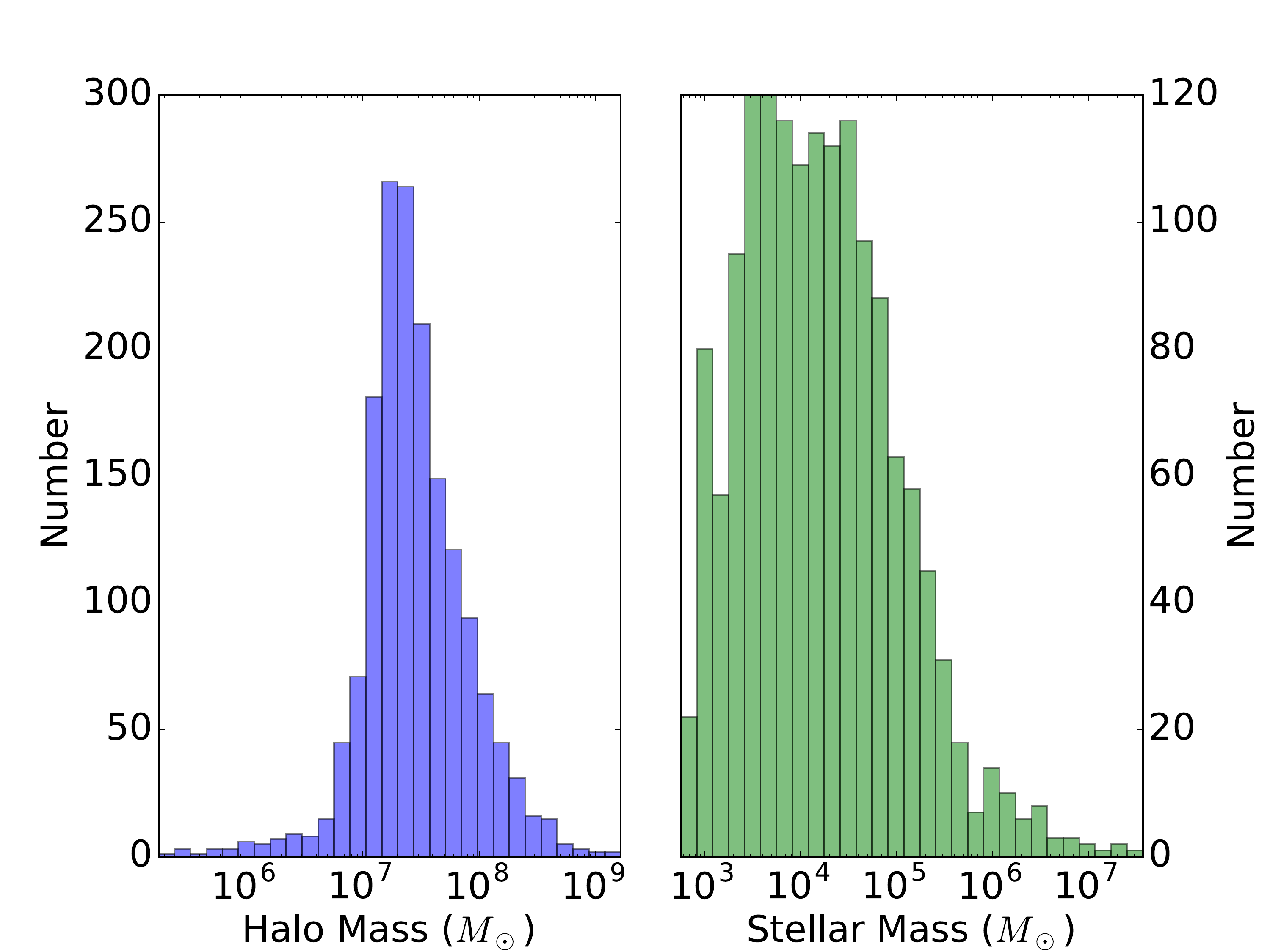}
\caption{Halo number counts in the Renaissance Simulations rare peak region for star-hosting haloes as a function of halo mass (left) and stellar mass (right).}
\label{fig:hist0}
\end{center}
\end{figure}

The Renaissance Simulations use individually run ``zoom-in'' subvolumes to produce an effective total resolution of $4096^3$ in a comoving box of size $(40~\rm{Mpc})^3$ using the hydrodynamic AMR code \enzo{}. The simulations are run using $\Omega_M = 0.266$, $\Omega_{\Lambda} = 0.734 $, $\Omega_b = 0.0449$, $h = 0.71$, $\sigma_8 = 0.8344$, and $n = 0.9624$ from the 7-year WMAP results \citep{2011ApJS..192...16L} with standard definitions for each variable. We analyse results from the ``rare peak'' zoom-in region of the Renaissance simulations with a top grid dimension of $512^3$ centred about the Lagrangian volume of two $5 \times 10^{10}\ \rm{M_\odot}$ haloes at $z = 6$. Due to the large size of the simulation, a large sample of 1654 galaxies are available for our analysis with an effective dark matter resolution of $2.9 \times 10^4\ \rm{M_\odot}$ at $z = 15$ and a spatial resolution of 19 comoving parsecs.  In the rare-peak zoom-in region, the most massive halo has a total mass of $1.62 \times 10^9\ \rm{M_\odot}$ and a stellar mass of $1.55 \times 10^7\ \rm{M_\odot}$ at $z=15$. We provide histograms of total and stellar mass of all star-hosting haloes in Fig. \ref{fig:hist0}. The Renaissance Simulations capture the formation of haloes and their star formation above $3 \times 10^6\ \rm{M_\odot}$ with more than 100 particles, resulting in an incomplete halo sample below this mass \citep{2013ApJ...773...83X,2016ApJ...823..140X}.  However, haloes start to form metal-enriched stars above $10^7\ \rm{M_\odot}$, as shown in the left panel of Fig. \ref{fig:hist0}, well above this resolution limit.  Thus, the drop off in number counts at low mass is caused by physical feedback effects, not lack of resolution.

\subsubsection{Star Formation Model}

We use the \citet{2007ApJ...659L..87A} Population III star formation and the \citet{2009ApJ...693..984W} radiative stellar cluster  routines to simulate the effect of star formation, radiative  and supernovae feedback on the cosmological environment.

We trigger Population III star formation within a converging gas flow when the $\rm{H_2}$ fraction exceeds $10^{-4}$, the metallicity fraction is below $10^{-4}\ \rm{Z_\odot}$, and the baryon overdensity exceeds $5 \times 10^5$. Population III stars are assigned a random mass according to the distribution 
\begin{equation}
  f(\log M)dM = M^{-1.3} \exp\left[-\left(\frac{40\ {\rm M}_\odot}{M}\right)^{1.6}\right] dM.
\end{equation}
Furthermore, we set the minimum mass for Population III stars to $5\ \rm{M_\odot}$ and limit the maximum mass to $300\ \rm{M_\odot}$.


Whereas the Population III star formation routine produces star particles representing individual stars, it is currently computationally prohibitive to model every metal-enriched star in a cosmological simulation so we employ a radiative stellar cluster routine. For gas with a metallicity above above $10^{-4}\ Z_\odot$, we trigger the creation of a star particle once a collapsing region has the properties of a molecular cloud, having a density corresponding to a dynamical time of $3\times10^6\ \rm{yr}$ ($\sim5 \times 10^{-22}\ \rm{g\ cm^{-3}}$) and would form a particle with a minimum mass of 1000\ $\rm{M_\odot}$ assuming that it inherits 7$\%$ of the cold gas in the star-forming cloud. Each particle therefore represents an entire star cluster with a distribution of individual stars.

In both routines, each star particle inherits the metallicity of its accreted gas and injects metals into the surrounding medium after a supernova. The Population III model triggers a pair-instability supernova for stars with masses between 140 and 260 $\rm{M_\odot}$ and a Type II supernovae for stars between 11 and 40 $\rm{M_\odot}$ \citep{2012MNRAS.427..311W}  at the end of a mass-dependent lifetime \citep{2002A&A...382...28S}. For metal-enriched stars, we inject
 $6.8 \times 10^{48}\ \rm{erg\ M_\odot^{-1}}$ over the 20 Myr lifetimes of the star particle \citep{1986ARA&A..24..205W} and initialize the supernovae in a resolved blast wave of radius 10~pc.

\subsubsection{Radiative Transfer and Chemistry Model}

To include the effects of photoionization and photo-heating, we solve the radiative transfer equation using \moray{} \citep{2011MNRAS.414.3458W}. We apply a time-dependent Lyman-Werner radiation background \citep{2005ApJ...629..615W,2012MNRAS.427..311W} and we use the star particles created from our star formation routines as point sources of Lyman-Werner and ionizing radiation. To reduce the computational load of ray-tracing individual photons throughout the simulation volume, we use an adaptive ray-tracing \citep{2002MNRAS.330L..53A} scheme applying the Hierarchical Equal Area isoLatitude Pixelation ({\sc HEALPix}) \citep{2005ApJ...622..759G} of a sphere to split photons as they travel from the source. This scheme subdivides  the surface of a sphere into 12 equal area pixels that can be further divided by factors of 4 resulting in $12 \times 4^{\rm{level}}$ equal area pixels. Photons travel along the normal of the {\sc HEALPix} pixels until they are split into higher {\sc HEALPix} levels when the number of photons per cell falls below 5.1.

For metal-enriched star clusters, we assume each star particle generates $1.12 \times 10^{46}\ \rm{erg\ M_\odot^{-1}}$ of ionizing radiation at 21.6~eV in a monochromatic spectrum. This implicitly assumes that radiating stellar clusters follow a Salpeter IMF with mass cut offs at 0.1 and 100 $\rm{M_\odot}$. Population III star particles are given ionizing luminosities consistent with their mass \citep{2002A&A...382...28S}. We also solve for the abundances of nine species (\hi{}, \hii{}, \hei{}, \heii{}, \heiii{}, $\rm{e^-}$, $\rm{H^-}$, $\rm{H_{2}^+}$, $\rm{H_{2}}$) with a non-equilibrium solver \citep{1997NewA....2..209A} that uses the chemical energy of each cell and a look-up table for metal cooling rates \citep{2008MNRAS.385.1443S} produced by the photoionization solver {\sc Cloudy} \citep{2013RMxAA..49..137F}.

\subsection{Halo Analysis}

We use the dark matter halo-finding code {\sc Rockstar} \citep{2013ApJ...762..109B} to generate a master list of self-gravitating bound regions. We then use {\sc Consistent Trees} \citep{2013ApJ...763...18B} to produce a halo merger tree.  We then further process the trees using \yt{} \citep{2011ApJS..192....9T} and a merger tree organizing code written for this project to create a time-ordered list of position and radii during halo assembly. 

Because self-gravitating regions may exist inside of larger haloes, our code takes the raw merger trees and removes these sub-haloes by treating the encompassing region as a single halo with the largest virial radius of the mutually bound haloes to create a new halo list. Using the new list, our code reconstructs the merger tree by assuming haloes are merged once their center falls within the virial radius of a largest of the merging haloes. The tree then continues from this halo after a merger. Using this method, the mass of a halo is calculated as the total mass of dark matter, gas, and star particles within a sphere with a radius equal to the virial radius taken from Rockstar and may include mass from more than one halo from the original tree.  

\subsection{Spectrum Building}
\label{Spectum}
Having modelled stellar clusters, metals, temperature, gas, and radiation throughout our simulation, we are left with a rich source of cosmological data with which to generate synthetic spectra. In our model, we treat each star particle as a star cluster consisting of a probabilistic distribution of radiation in frequency space based on models using the metallicity and age-dependent Flexible Stellar Population Synthesis code \citep[FSPS;][]{2010ApJ...712..833C} which is tuned to be consistent with observations. We model our particles assuming an aggregated mean of observed stars of similar composition and age which is consistent with the probabilistic cluster model used to create metal-enriched star particles and model radiative transfer in the cosmological simulation. This allows us to enforce consistency between stellar feedback in the cosmological simulation and stellar feedback in radiative transfer post-processing. However we note that the mean ionizing energy of photons from stellar population synthesis will be slightly different than the energy assumed in the simulation as the stellar population evolves from brighter stars with more ionizing photons to dimmer, longer-lived stars with a redder spectra as shown in Fig. \ref{fig:ccplot} left. We expect this to cause slightly larger than physical ionized regions around older stars in the simulation, but this is tempered somewhat by the overall youth of stellar populations at $z =  15$.

\subsubsection{Dust and Gas Model}
\label{sec:DustGas}

We use the Monte Carlo ray-tracing code {\sc Hyperion} \citep{2011A&A...536A..79R} to propagate, extinguish, and scatter photons through a dusty medium. To initialize the radiation source, we arrange each star particle within a 3-D grid and apply the bolometric luminosity and spectra calculated with {\sc FSPS}. We use the full grid hierarchy from the \enzo{} output to create a derived equivalent AMR grid with a predefined maximum highest level. Due to purely computational limitations, we select the highest grid level that will produce no more than $45^3$ cells. Relevant physical quantities are then applied to the derived grid from the corresponding cells in the \enzo{} output.

For gas extinction, we use {\sc Cloudy} to produce an isotropic frequency and density dependent hydrogen opacity relationship and allow {\sc Hyperion}'s intrinsic local thermodynamic equilibrium tool to extrapolate emissivities over a broad specific energy range. We apply this model to the density of neutral hydrogen in each cell of our grid and assume that neutral hydrogen accounts for the greatest component of extinction of the nine species calculated in the cosmological simulation. For dust extinction, we use the  \citet{2003ARA&A..41..241D} ($\rm{R_v} = 2.1$) model and assume that dust contributes $7\%$ of the mass of metals in a cell. We also assume a solar abundance pattern and weight the metal density of each cell accordingly to produce a dust density. Our model uses $10^8$ photons to produce a resultant spectra of 8,000 frequencies from 0.05 to 5 microns.  We limit the photon propagation to the virial radius of the galactic halo and assume an optically thin medium to the observer outside of the halo.

\subsubsection{Emission Lines}

We also consider nebular emission lines from the chemical network in the warm ($T>7000$~K) and ionized regions in the vicinity of stellar clusters. While we track nine primordial species and a metallicity field to approximate the ISM and accurately model galaxy formation in the simulation, a more refined model is required to get the full complement of metal emission lines. 

Once again, we turn to {\sc Cloudy}'s photoionization solvers. To generate a high-resolution region for analysis, we reuse the derived AMR grid used for {\sc Hyperion} to extract gas properties. Using the full spectra from {\sc FSPS} for each star particle, we sum the spectra and luminosity within each cell and treat the sum as a single source in the cell center. 

We use {\sc Cloudy} to solve and simulate the full complement of emission lines within each cell, constraining the model inputs to the spectra and luminosity from {\sc FSPS}, the hydrogen density, and the metallicity using a solar abundance pattern \citep{2009ARA&A..47..481A} from the cosmological simulation to try to keep our photoionization calculation as close to our cosmological results as possible. Because of the dynamic, non-equilibrium nature of cosmological \hii{} regions, we stop the simulation once the electron fraction reaches a value consistent with the local mean electron fraction from the \enzo{} chemistry solver rather than let {\sc Cloudy} find its own equilibrium. The emission lines from each cell are then summed to produce a single source emission line distribution.

We found that the relative line strengths were mostly invariant for a range of inner gas radii centred around $10^{17}$~cm within a sample of large haloes with a large number of individual star particles. We therefore assume the gas to have a minimum radius of $10^{17}$ cm (proper) and maximum radius equal to the half-width of the cell for each calculation of the emission emerging from each cell. Stellar clusters have a dynamic and complicated radial luminosity distribution, but our cosmological simulation only captures entire clusters for metal-enriched stars so we acknowledge the limitation of this treatment. 

Finally, we apply a frequency dependent ratio of the {\sc Hyperion} resultant flux to the {\sc FSPS} intrinsic flux to the corrected {\sc Cloudy} emission lines consistent with \citet{2015MNRAS.454..269P} to simulate extinction and scattering of lines. We find that this method is more feasible than attempting to use enough photons to simultaneously resolve both emission lines and continuum with {\sc Hyperion}. Our final spectra includes the sum of the {\sc Cloudy} results for each sub-region and the {\sc Hyperion} results for the entire halo.

\subsection{Filters, Magnitudes, and Images}

\subsubsection{Filters}

Our analysis includes comparisons between our synthetic spectra as seen through HST and JWST by applying filters to our results. For Hubble, we use filters from WFC3 used for the later HUDF surveys. For JWST we assume the use of the Near-Infrared Spectrograph (NIRSpec) and the Near-Infrared Camera (NIRCam). Additionally, we apply a redshift to our spectrum to allow for direct comparison in the observer's frame. We report the total bolometric luminosity and intensity for telescope filters assuming an optically thin path to the observer outside of the virial radii unless otherwise stated.

\subsubsection{Filter Fluxes}

\begin{figure}
\begin{center}
\includegraphics[scale=.35]{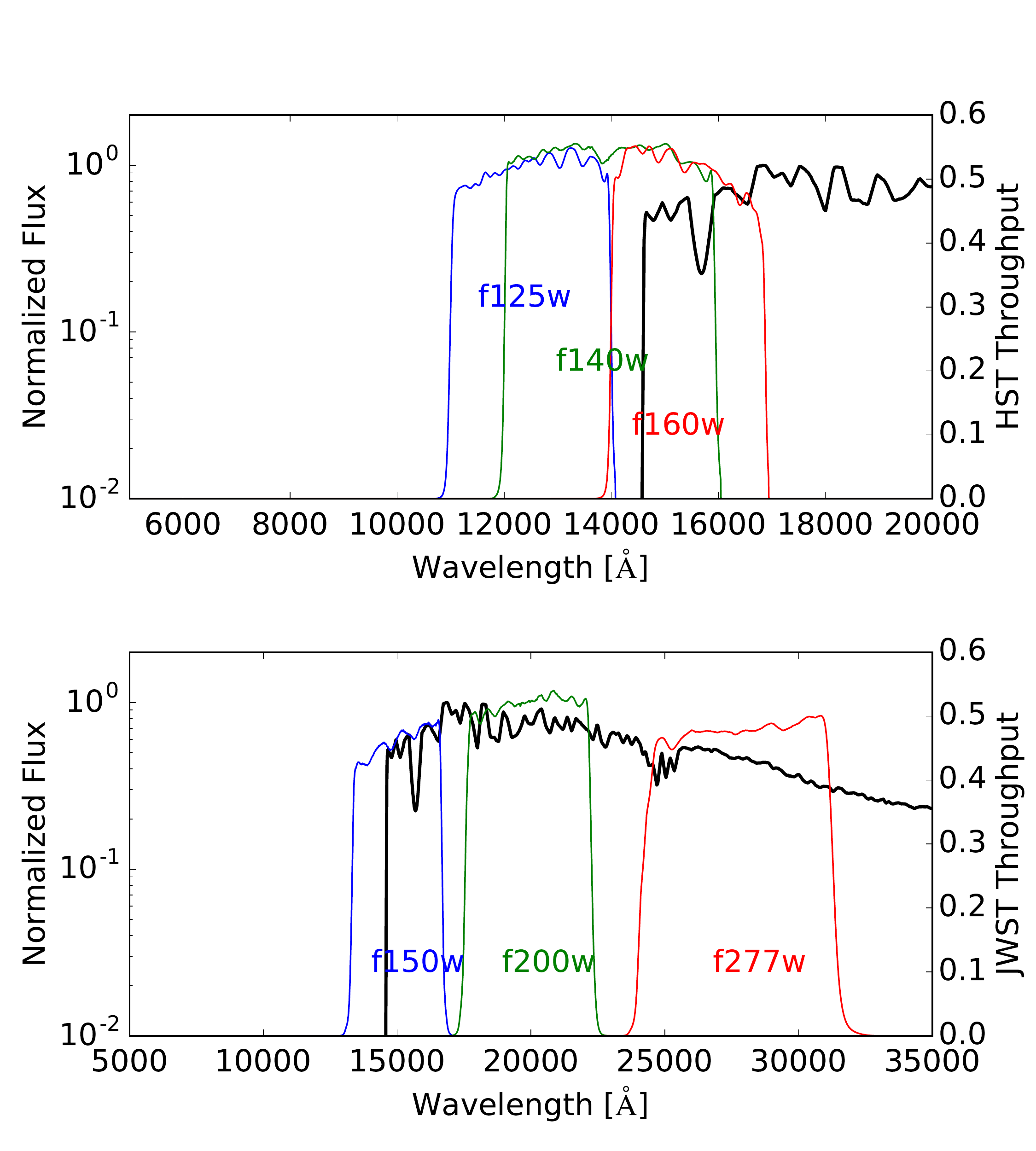}
\caption{A sample synthetic spectra placed at $z = 15$ overlaid with selected HST and JWST filters. From top to bottom: A plot of the overlapping HST WFC3 wide band IR filter throughputs and a plot of JWST NIRCam wide band IR filter throughput, normalized to the largest value in units of erg s$^{-1}$ cm$^{-2}$ \AA$^{-1}$.}
\label{fig:filters0}
\end{center}
\end{figure}

We also calculate the absolute and apparent magnitudes in each filter. Filter throughputs in the observed frame are plotted against a sample spectra of an object at $z = 15$ in Fig. \ref{fig:filters0}. We eliminate the need for a $K$-correction by transforming the filter responses into the rest frame when integrating the spectra \citep{1968ApJ...154...21O}. Therefore, we use a sum
\begin{equation}
\label{eq:fluxexpect}
f(\nu_{\rm{o}}) =  \frac{1}{4 \rm{\pi} d_{\rm{L}}^2} \int \frac{L_{\nu}(\nu_{\rm{e}})}{\nu_{\rm{e}}} R(\nu_e) d\nu_{\rm{e}},
\end{equation}
to estimate the flux over the passband where $\nu_{\rm{e}}$ and $\nu_{\rm{o}}$ are the wavelength in the rest and observed frame respectively, $R(\nu_e)$ is the transformed filter throughput, and $d_{\rm{L}}$ is the luminosity distance at a redshift $z$.

\subsubsection{Synthetic Imaging}

Two dimensional images from {\sc Hyperion} are produced using the built-in method of binning photons from the Monte Carlo gas extinction process at several predefined individual wavelengths. Images produced using this method do not include emission lines produced in nebulae, but include dust and gas scattering and emission as described in Section \ref{sec:DustGas}. To produce synthetic observations of early galaxies viewed at the present day, we process the raw luminosity of each pixel to account for several distance, resolution, noise, and blurring effects.

Our monochromatic flux per pixel is given simply as $F = L(1+z)/(4\pi d_{\rm{L}}^2)$ presented in units of $\rm{erg\ s^{-1}\ cm^{-2}\ Hz^{-1}}$. To account for the observed resolution through an expanding cosmology, we take the angular distance to be $d_{\rm{A}} = d_{\rm{L}}/(1+z)^2$. For a given resolution and flat universe, the proper width of a pixel is therefore given by 
\begin{equation}
W_{\rm{pixel}} = d_{\rm{A}} \theta = \frac{c \theta}{H_{\rm{0}}(1+z)}\int_0^z \frac{dz'}{\sqrt{\Omega_{\rm{M,0}} (1+z')^3 + \Omega_{\rm{\Lambda,0}}}}.
\label{eq:res}
\end{equation}
Here $\theta$ is resolution of the space telescope camera. We use resolutions of 0.065\arcsec{} for NIRcam and 0.15\arcsec{} for WFC3 to resize the native simulated image to the appropriate angular resolution for a given redshift. We apply two interpolation schemes to images to simulate instrument effects. First we apply a Gaussian blur with a standard deviation of five instrument pixels to the image. We then resize the the image to the instrument pixel size and applying pixel-binning interpolation to produce the final image. For the purpose of comparison, we provide images in the native resolution of the simulation and the same images after processing.

To account for noise, we assume JWST and Hubble sensitivities of $\sim 10^{-8}\ \rm{Jy}$ and $\sim 2.5 \times 10^{-8}\ \rm{Jy}$ respectively with a $S/N = 10$ after an exposure time of $10^4$~s at the infrared frequencies of interest. We then extrapolate to a 1~Ms ($\sim 11.6$ days) exposure for a final noise value of $\sim 10^{-11}\ \rm{Jy}$. Assuming a Gaussian distribution for noise, we take the mean and standard deviation to be half that value and add the noise flux directly to the image after resizing the pixels.

\begin{figure}
\begin{center}
\includegraphics[scale=.35]{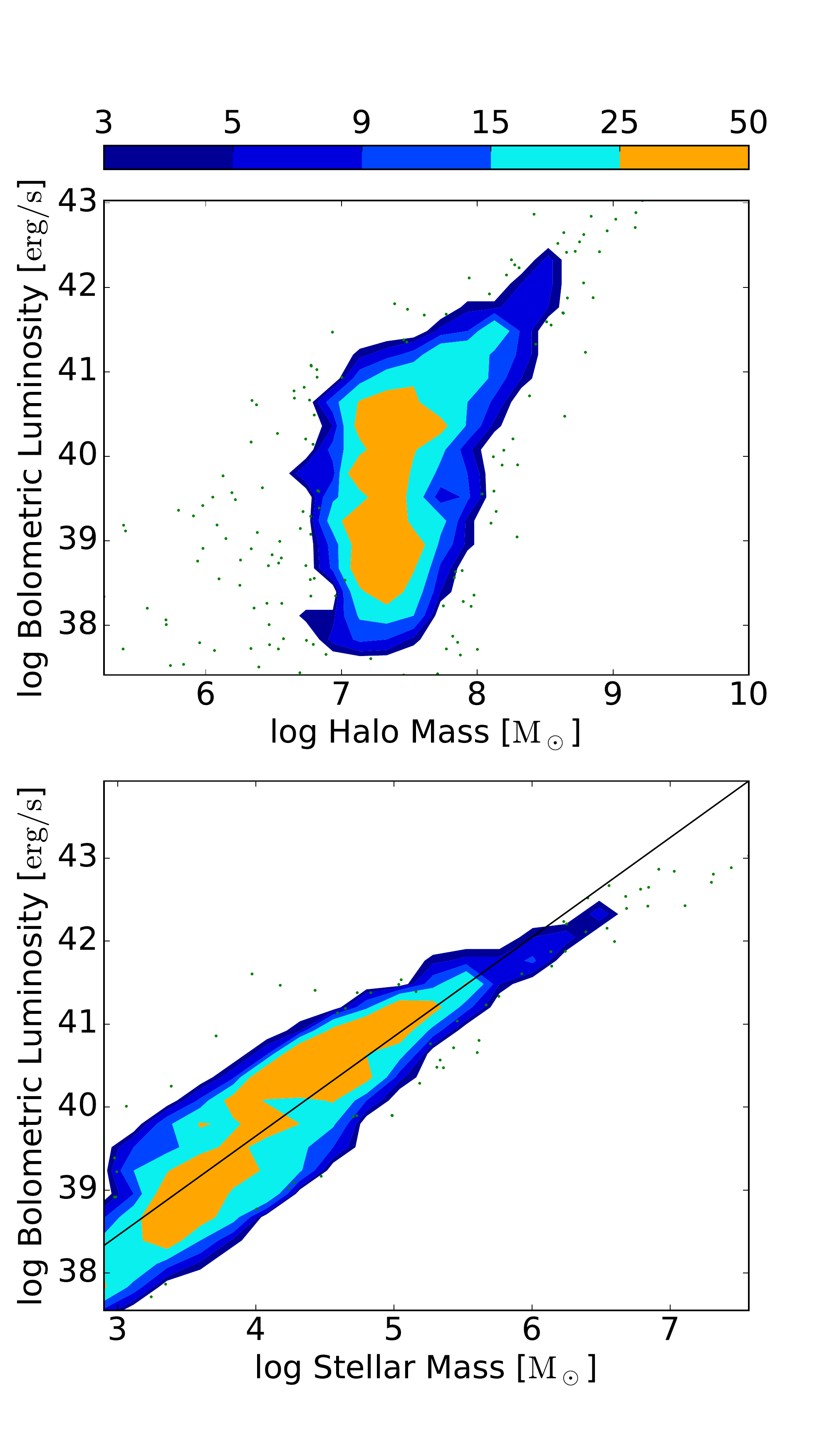}
\caption{Contour histograms and scatter plots of total bolometric luminosity versus halo mass (top) and stellar mass (bottom), showing a wide distribution of luminosities for the haloes with masses below $10^8 \Ms$, however a clear relationship between stellar mass and luminosity is present with the luminosity scatter caused by differing bursty star formation histories. The fit for the bolometric luminosity regression is presented in Table \ref{tab:slopeshubble}.}
\label{fig:lum1}
\end{center}
\end{figure}

\begin{figure*}
\begin{center}
\includegraphics[width=0.46\textwidth]{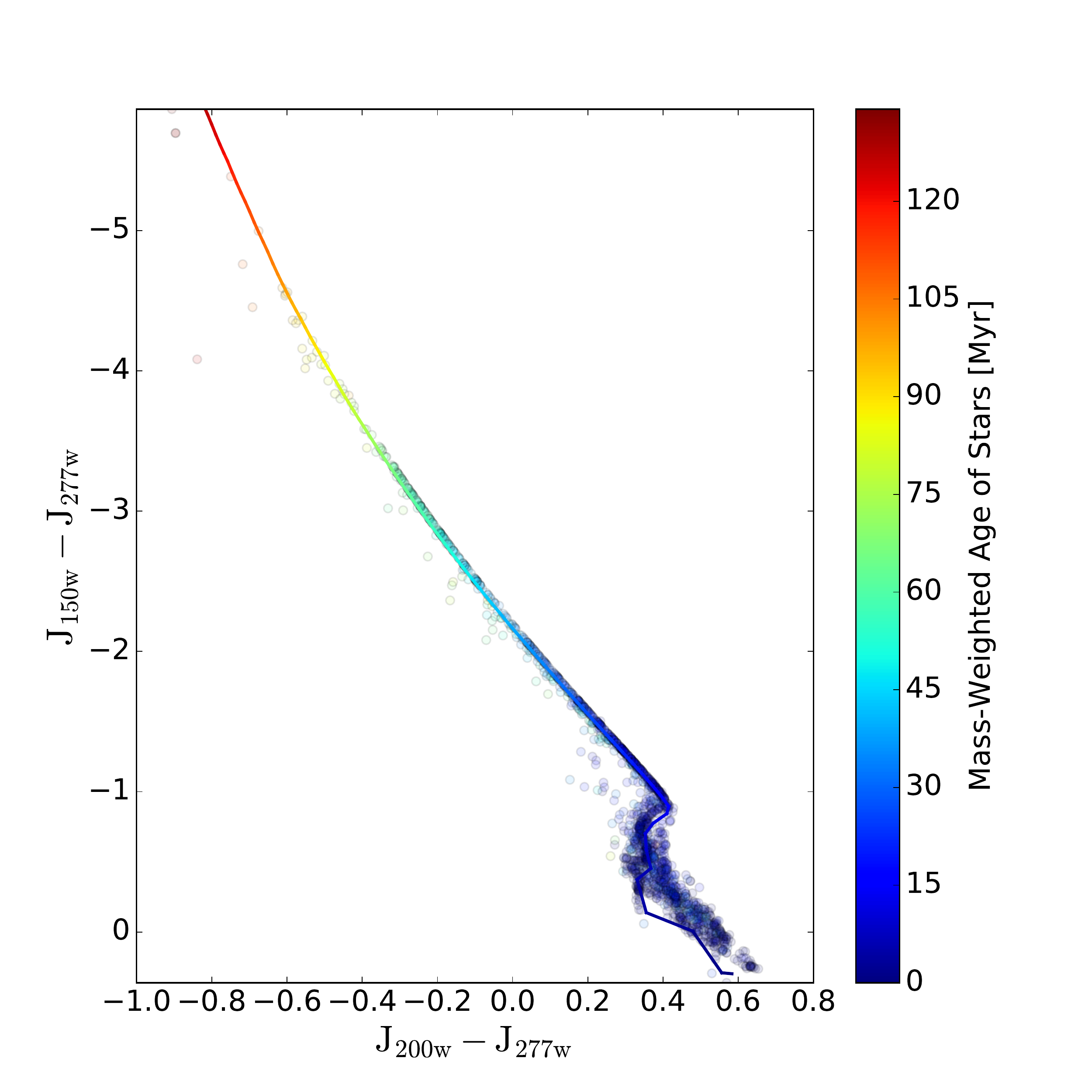}
\hfill
\includegraphics[width=0.46\textwidth]{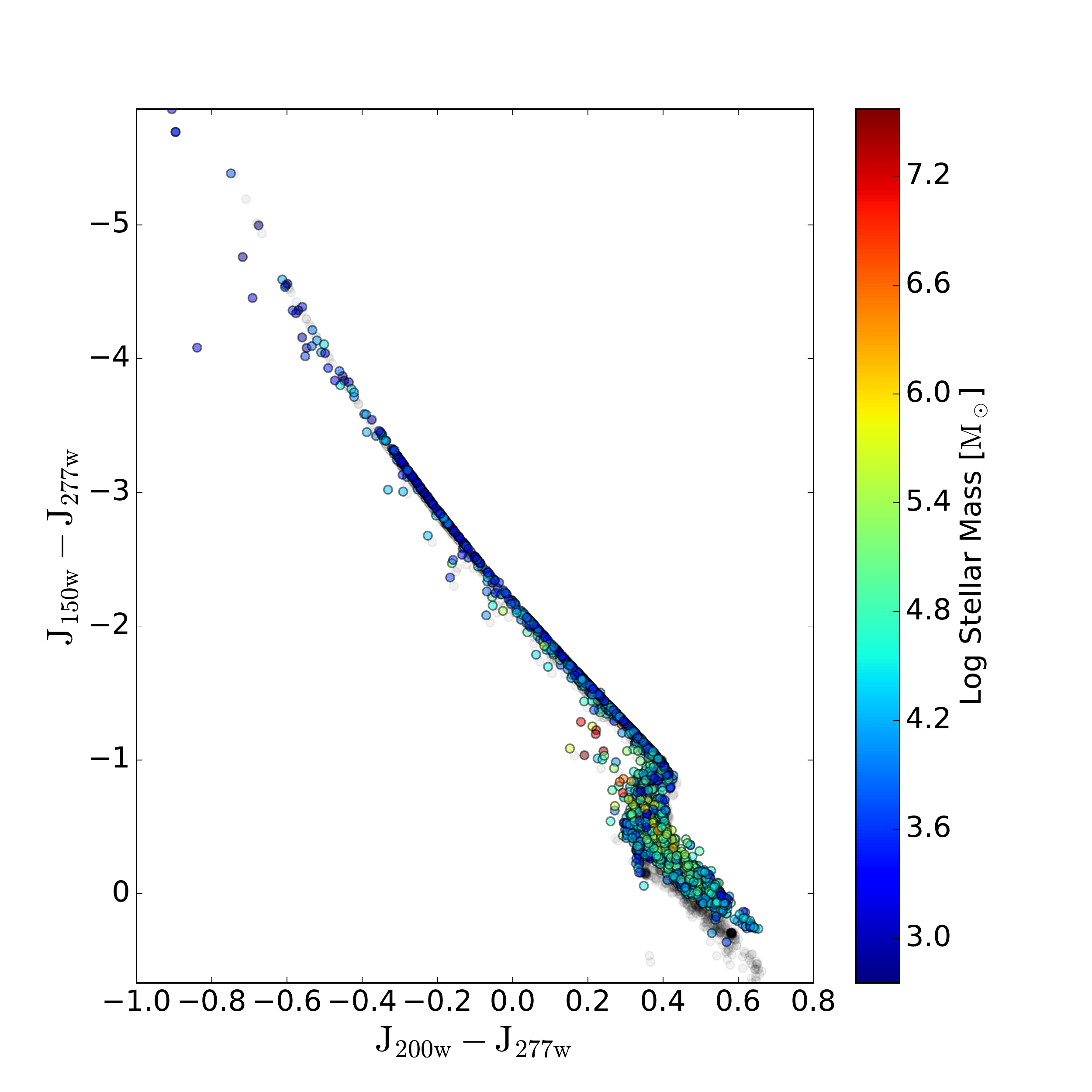}
\caption{Colour-colour plots of star-hosting haloes coloured by their  mass-averaged stellar age (left) and stellar masses (right), when the galaxy spectra are redshifted to $z = 15$. The line in the left plot represents the evolution of a single 0.01 $\rm{Z_\odot}$ stellar cluster coloured by the same range of ages. The grey points right plot represent unprocessed stellar colours which mostly overlap the processed JWST colours with a small offset. Most of the least massive young galaxies lie in a single line as their small stellar populations reliably follow the line, whereas more massive galaxies congregate near J$_{200w}$ - J$_{277w}$ $ \sim 0.25$, having a stellar population composed of stars with a variety of ages and metallicities.  The mean stellar age plays the biggest role in determining the typical galaxy colours, given the bursty nature of star formation in these high redshift galaxies.}
\label{fig:ccplot}
\end{center}
\end{figure*}

Since we processed a fixed number of monochromatic images in our initial analysis, the redshifts we present in our images are fixed by the average wavelengths of each filter, $\rm{z} = \langle\lambda_{\rm{filter}}\rangle/\lambda_{\rm{image}} -1$. This results in two to four redshifts in each filter band based on our initial sample of wavelengths. We choose 1500~\AA(rest) as the wavelength for both the Hubble and JWST filter images to avoid proximity to strong emission lines. We note that this implicitly assumes a similar star formation history for haloes at $8 \la z \la 15$ and so we perform this procedure as an exercise with this understanding.

To produce optical composite images, photons are binned at 50 wavelengths between 3800 and 7500 \AA(rest). We use a multi-lobe piecewise Gaussian fit of the CIE XYZ colour matching functions \citep{Wyman2013xyz} for the flux in each pixel and then transform to RGB using the CIE E white point matrix. The intensity of each pixel in the image is then scaled to the power of 1/4 to accentuate light scattering.

\begin{figure}
\begin{center}
\includegraphics[scale=.33]{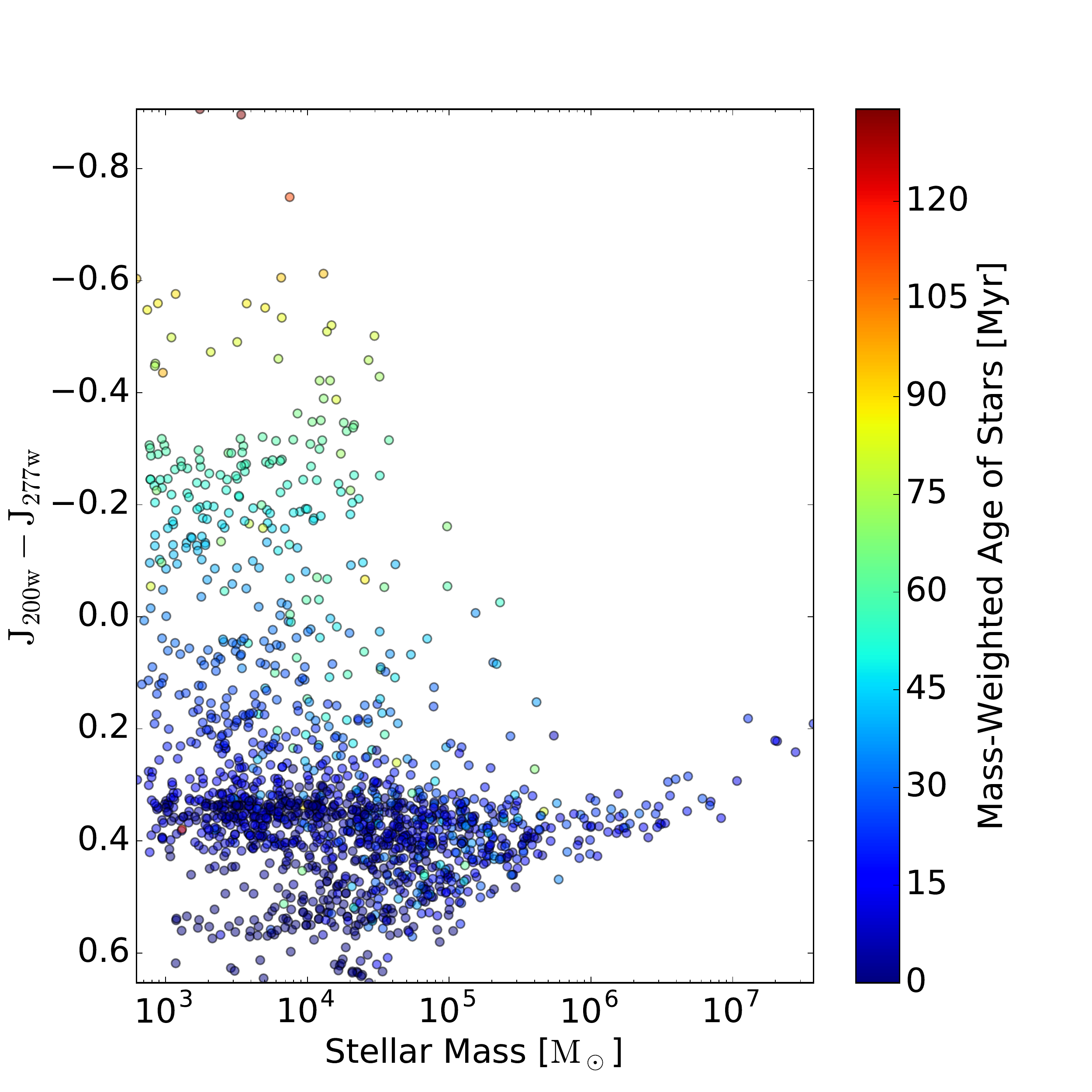}
\caption{Colour J$_{200w}$ - J$_{277w}$ versus stellar mass coloured by the mass-weighted stellar age, assuming $z=15$, showing a clear trend toward older populations in the least massive galaxies, whereas galaxies with $M_\star \gsim 10^6 \Ms$ show a slight decreasing trend with increasing stellar mass.}
\label{fig:greenvalley}
\end{center}

\end{figure}

\begin{figure}
\begin{center}
\includegraphics[scale=.33]{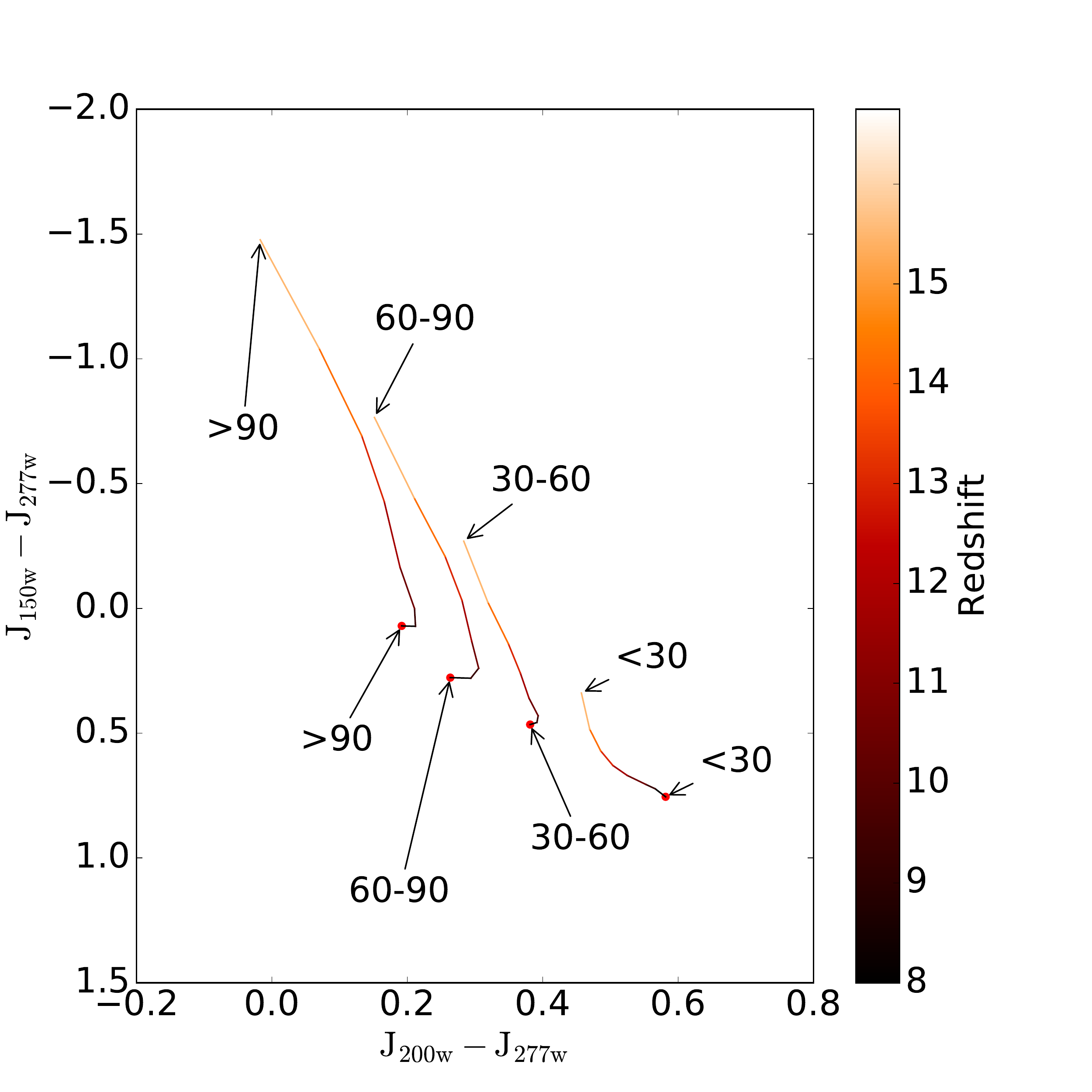}
\caption{The evolution of average colour-colour plot as the spectra are redshifted in the range $z=8-15$ when categorized by mean stellar age in the ranges (in units of Myr) indicated in the plot.  The Lyman break causes the reddening in J$_{150w}$ -- J$_{277w}$ with redshift, whereas the older stellar populations trend redward in J$_{200w}$ -- J$_{277w}$.}
\label{fig:zevo}
\end{center}
\end{figure}

\begin{figure*}
\begin{center}
\includegraphics[scale=.37]{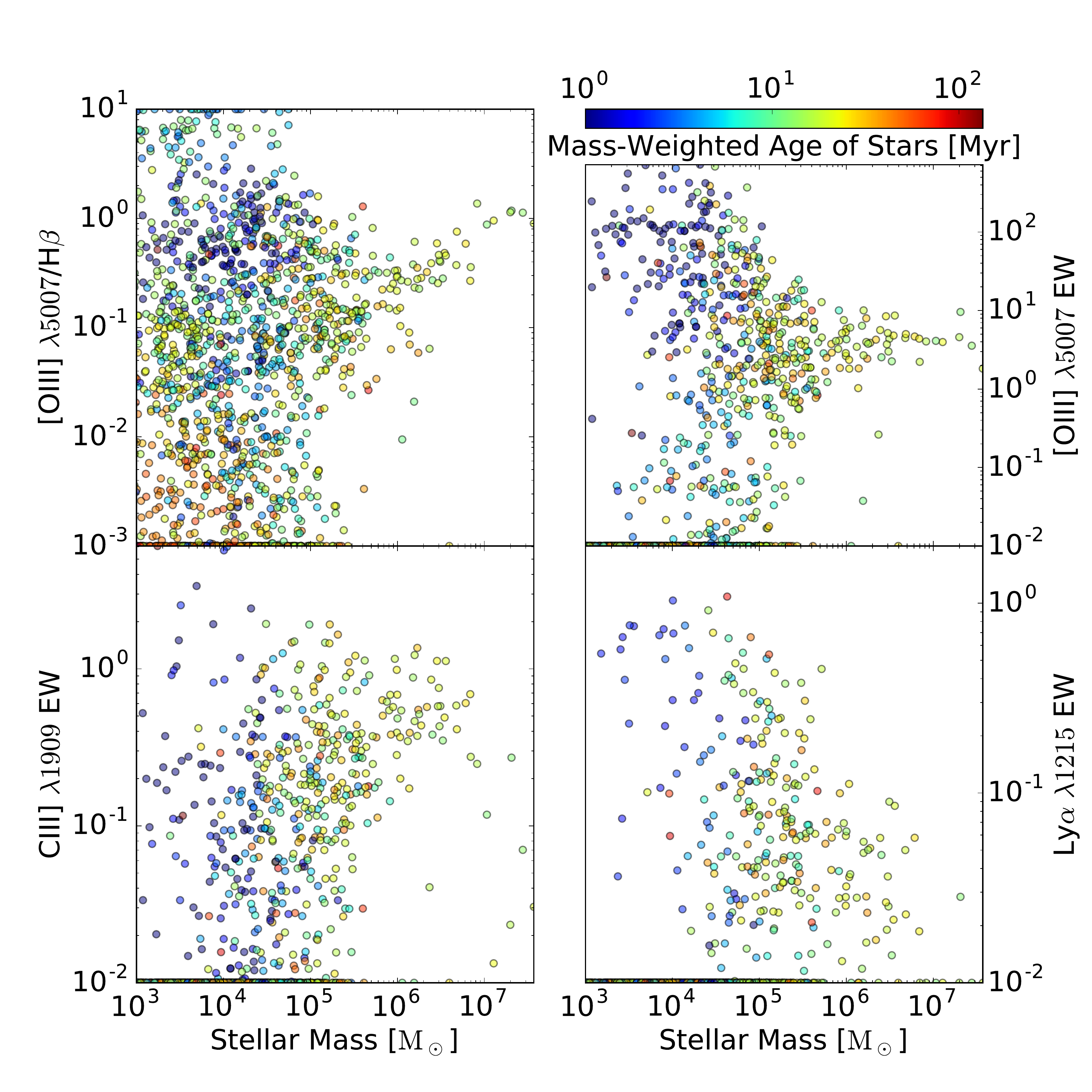}
\caption{Emission line measures as a function of stellar mass.  Upper left: \oiii{} $\lambda$5007 to H$\beta$ ratio.  Upper right: \oiii{} $\lambda$5007 equivalent width (EW). Bottom left: \ciii{} $\lambda$1909.  Bottom right: Ly$\alpha$ equivalent width. All emission line equivalent width and ratio plots are coloured by mass-weighted average age of the stellar population with definitions in the colour bar. Extreme values at low mass were plotted at the extreme of the plot window.}
\label{fig:lines0}
\end{center}

\end{figure*}

\section{Results}
\label{AggHalo}

We segregate our results into a section on the aggregate statistics on our entire sample and a section on synthetic observations of two individual large galaxies. For the entire star-containing halo population, we provide luminosity curves, colour-colour plots, emission line ratios and equivalent widths, composite spectra, and  BPT diagrams and correlations with stellar mass. For the individual galaxies, we produce synthetic images as seen by HST and JWST and compare to the physical properties of the dark matter halo and the associated galaxy. We also compare weighted mean values for each of these parameters to total flux with respect to viewing angle for the largest galaxy in our simulation.

\subsection{Aggregate Halo Statistics}

Our sample of 1654 galaxies allow us to explore the distribution of spectra, emission lines, and luminosity of haloes with respect to their mass and composition. Fig. \ref{fig:lum1} shows the bolometric luminosity distribution of our sample with respect to halo and stellar masses. haloes between $10^{7}$ and $10^{8}\ \rm{M_\odot}$ form the boundary between star containing galaxies and non-luminous haloes due to both the inefficient cooling inhibiting star formation and shallower potential wells. This boundary appears on the plot as a $\sim 10^4$ range in luminosity inside this mass range. Higher halo masses are much more strongly correlated to luminosity, but our sample size diminishes at this end of our mass distribution. The relationship between luminosity and stellar mass is much more orderly and apparently linear, however luminosities still vary as much as a factor of 30  between $10^{3}$ and $10^{4}\ \rm{M_{\odot}}$. This is mainly due to variations in the composition and distribution of stars within each halo.  
Because haloes hosting small stellar populations are typically low mass, they are prone to photoevaporation and gas blowout \citep{2014ApJ...795..144C} by supernovae which can occasionally  leave stellar populations exposed, cutting off star formation. Galaxies with low stellar masses are therefore found to have a wide range of mean stellar ages consisting of either populations of bright, young stars or dim, older stars. At high stellar mass, the variability in luminosity is reduced to less than order of magnitude for a given stellar mass. In deeper potential wells, much of the remaining variability comes from the recent formation history of the galaxy. Systems with high merger rates or recent major mergers have disrupted gas distributions and bursts of star formation which may increase their luminosity with respect to stable, more isolated galaxies. These galaxies are also prone to stronger outflows due to the increased supernovae rate associated with higher star formation rates, but this is moderated by the deepness of their potential wells and the density of their gas.

The left panel of Fig. \ref{fig:ccplot} shows a colour-colour plot of our sample of galaxies coloured by mass-weighted mean stellar age. We observe a roughly linear relationship in magnitude space with a tendency for haloes with older stellar populations to inhabit a tail from $\rm{J_{150w}-J_{277w}} < -1.0$ and $\rm{J_{200w}-J_{277w}} < 0.4$. A well-correlated ``ridge" forms in the figure as the age of the mean stellar population increases, moving the peak emission out of the UV and into less variable portions of the spectra. These haloes are usually the low to intermediate mass systems that have lost enough gas to extinguish star formation. The redness of their spectrum can mostly be explained by an older stellar population dominated by stars with ages closer to 100 Myr. A line tracing the evolution of a single 1000 ${\rm M}_\odot$, 0.01 ${\rm Z}_\odot$ stellar cluster is shown to mostly follow the distribution with some notable exceptions redward of the line due both to the addition of emission lines and dust and gas attenuation. haloes with high stellar mass ($>10^6\ \rm{M_\odot}$) are seen to congregate in the centre of the plot around $\rm{J_{150w}-J_{277w}} = 0.5$ and $\rm{J_{200w}-J_{277w}} = 0.4$ (right panel of Fig. \ref{fig:ccplot}). Furthermore, the reddening of the galactic spectra is most apparent in larger galaxies with deeper potential wells and higher densities of gas and dust. When the processed colours are compared against the intrinsic FSPS composite, colours are seen to decrease by as much as 0.05 in $\rm{J_{200w}-J_{277w}}$ and 0.5 in $\rm{J_{150w}-J_{277w}}$. 

\subsubsection{Emission Line Strengths and Ratios}
\label{Sec:chemline}

Fig. \ref{fig:greenvalley} shows the JWST colour of each halo versus stellar mass. haloes with older stellar populations generally occupy the red portion of the plot (Low $\rm{J_{200w}-J_{277w}}$) as expected. haloes with higher stellar mass tend to have $\rm{J_{200w}-J_{277w}}$ values between 0.0 and -0.9 while intermediate and low stellar mass haloes appear to scatter without correlation between $\rm{J_{200w}-J_{277w}}$ values of 0.7 and 0.2. While the range can be explained by the source spectra and stellar age, some of the variability comes from processing through gas and dust.

By changing the filtering of the full spectra, we are able to recalculate colours for different redshifts with the assumption that the mean stellar age is an approximate proxy for a comparable star formation history if we translate the star formation rates from higher redshift to lower redshift for the duration of the halo's assembly. However while we note that this is usually only valid for extraordinary cases, as an exercise, the effect of an uneven sampling of our galactic SED due solely to red-shifting and mean stellar age is presented due to the significant impact it has on our final colours.  Fig. \ref{fig:zevo} shows the mean values of the colour-colour diagram in Fig. \ref{fig:ccplot} evolved from $z=8$ to $z=15$ for four bins of stellar age in 30 Myr increments. As redshift increases, 
filters move towards the top left of the plot as filters begin to overlap with the Lyman break. Older stellar populations were found to exhibit higher $\rm{J_{200w}-J_{277w}}$ and $\rm{J_{150w}-J_{277w}}$ at all redshifts  allowing the colour-colour plot to act as a tracer of both mean stellar age and redshift during reionization. When binned by log stellar mass, the data did not exhibit an observable difference in either colour at any individual redshift. 



We provide plots of the equivalent widths of the \oiii{} $\lambda$5007, \ciii{} $\lambda$1909 and Ly$\alpha$ lines for our entire sample in Fig. \ref{fig:lines0}. The relationship between halo stellar mass and equivalent width variance is seen to be generally inversely proportional to mass. Due to the extreme variance at low mass, we disregard un-observable equivalent widths below 0.01 \AA. We expect that these data result from haloes with extraordinarily low metallicity or gas mass ratios such as haloes that have photo-evaporated or haloes without young stellar populations.

An initial burst of star formation in small haloes produce higher \oiii{} EW and \oiii{}/H$\beta$ ratios as the halo is heated and ionized. Star formation is subsequently shut off since gas can no longer efficiently cool and supernovae expel most of the gas in the shallower haloes. haloes that are photo-evaporated by their proximity to larger galaxies, or are otherwise isolated from gas inflows, or have their star formation inhibited have progressively older stellar populations and drift downward in both \oiii{} plots. haloes that accrete enough cool gas to resume star formation until their haloes are again heated and ionized, restarting the cycle. haloes that continue to grow in this manner trace a ``zig-zag'' path as their stellar masses increase. Above a stellar mass of around $10^5\ \rm{M_\odot}$, haloes are typically large enough that star formation events create confined \hii{} regions and a single halo can consist of multiple regions of high and low formation rates. These haloes have lower variability in the \oiii{} EW and \oiii{}/H$\beta$ ratio plots as younger and older stellar populations are averaged. 

In galaxies with $M_\star \leq 10^{4.5}\ \rm{M_\odot}$, the highest \ciii{} EWs are correlated to young stellar populations but are still only 1--3~\AA. Above that mass, larger \ciii{} EWs are associated with intermediate age populations and galaxies with a stellar masses more than $\sim 3 \times 10^5 \rm{M_\odot}$. This behaviour is not dissimilar to the distribution of \oiii{} EW with respect to stellar mass and age, but exhibits more variability at high mass.

We show intrinsic rather than emergent Ly$\alpha$ equivalent widths, which would require the implementation of a dedicated Ly$\alpha$ radiative transfer code. We find that intrinsic Ly$\alpha$ equivalent widths are generally insubstantial and the maximum Ly$\alpha$ EW further decreases with increasing $M_\star$ from $\sim$ 1\AA\ for $M_\star \leq 10^5$ down to less than 0.1\AA\ for $M_\star \geq 10^7$ due to the lower neutral hydrogen fractions in bright galaxies.

\begin{figure}
\begin{center}
\includegraphics[scale=.35]{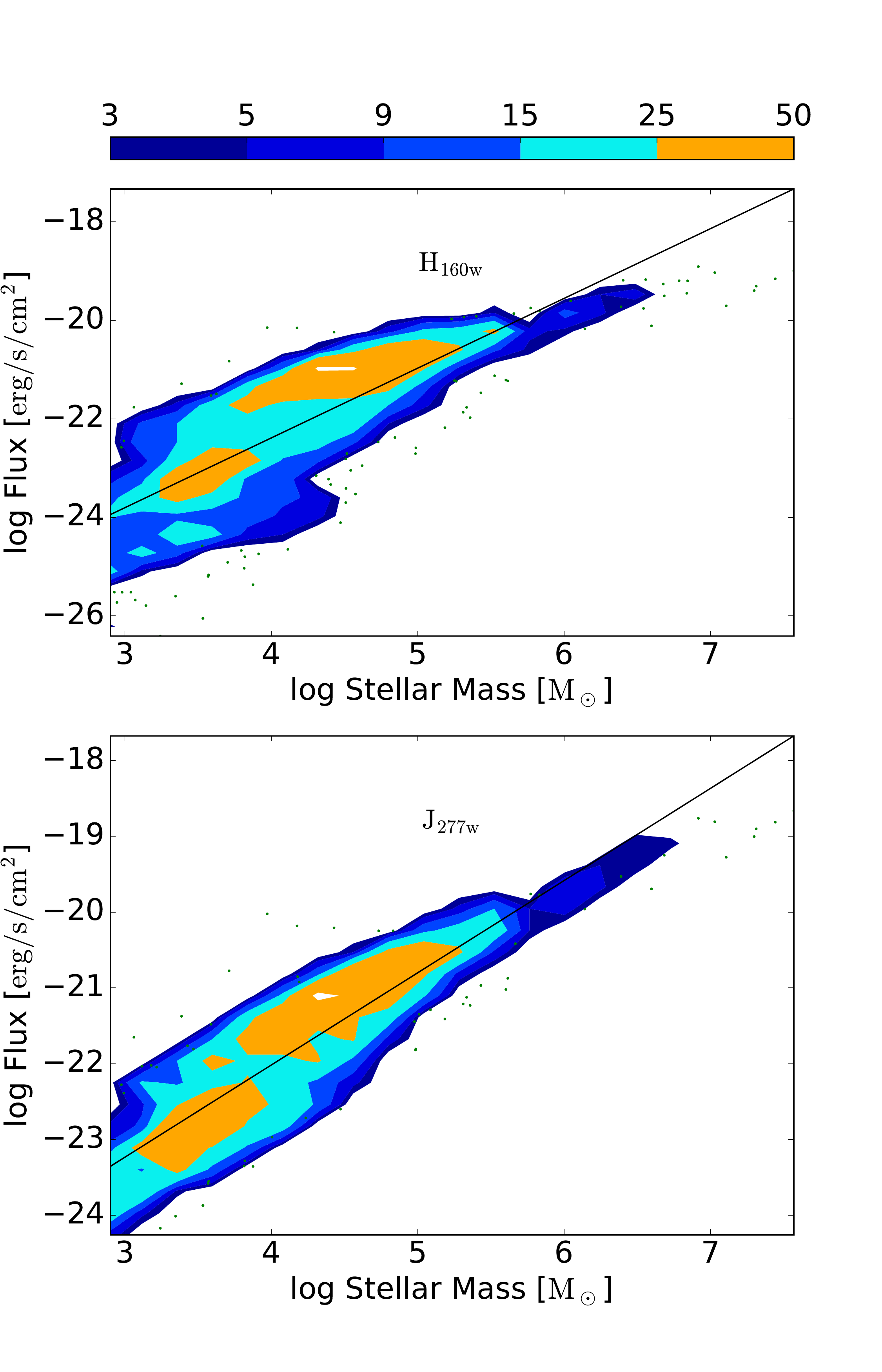}
\caption{Contour histograms and scatter plots corresponding to the Hubble f160w (top) and JWST f277w (bottom) filters with fits described in Table \ref{tab:slopeshubble}. While the regressions are appropriate at lower stellar masses, they tend to overestimate flux for log $M_\star$ > 6.}
\label{fig:lum2}
\end{center}
\end{figure}

\begin{figure*}
\begin{center}
\includegraphics[width=0.95\textwidth]{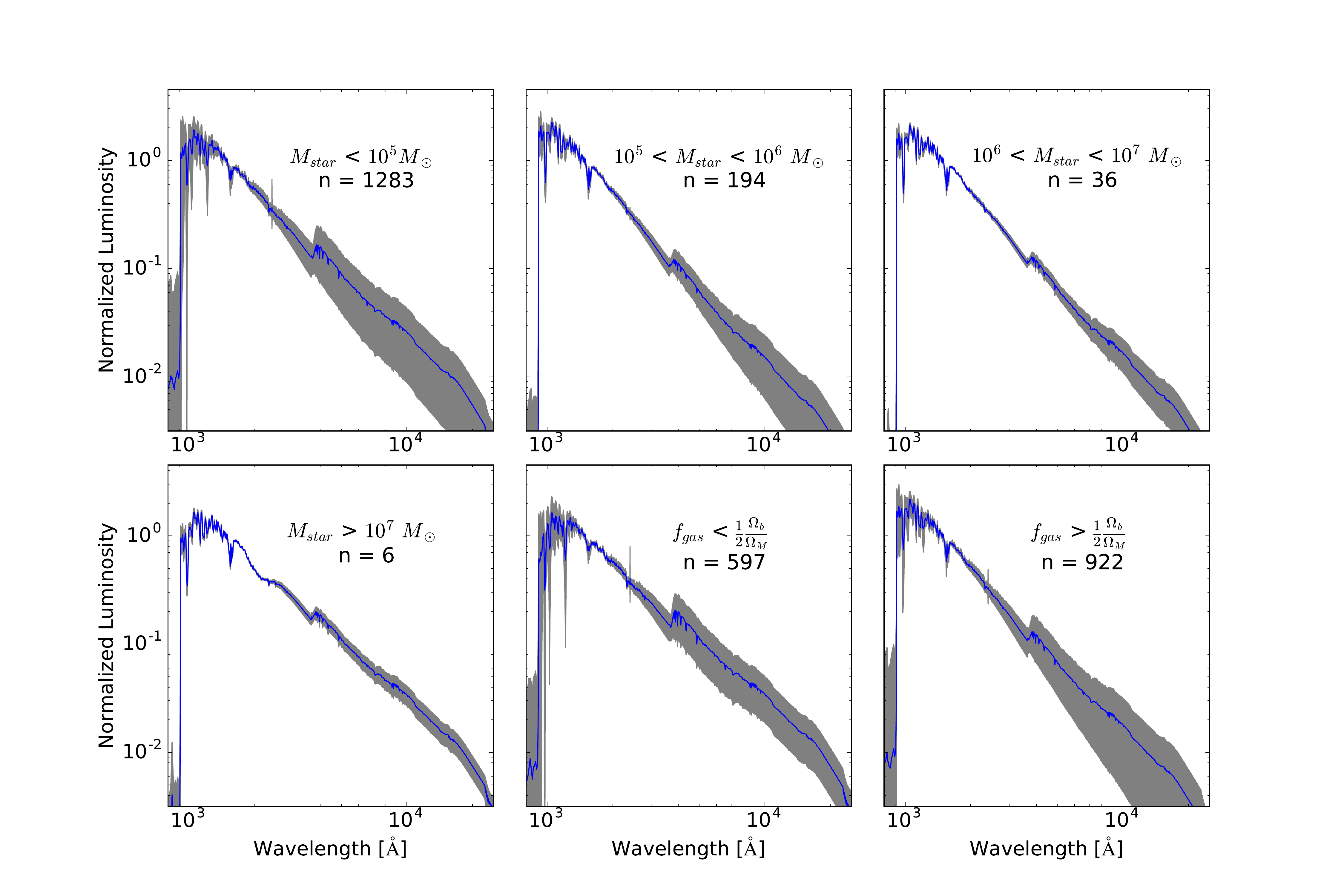}
\caption{Stacked galactic spectra with mean values in blue and 1$\sigma$ bands in grey in different stellar mass range (top row and bottom right) and gas-poor (bottom middle) and gas-rich (bottom right) haloes.  Luminosities are normalized to the mean value at 1500~\AA(rest), and the overall luminosities can be inferred from Fig. \ref{fig:lum1}.}
\label{fig:spectra0}
\end{center}

\end{figure*}

\begin{figure*}
\begin{center}
\includegraphics[width=0.75\textwidth]{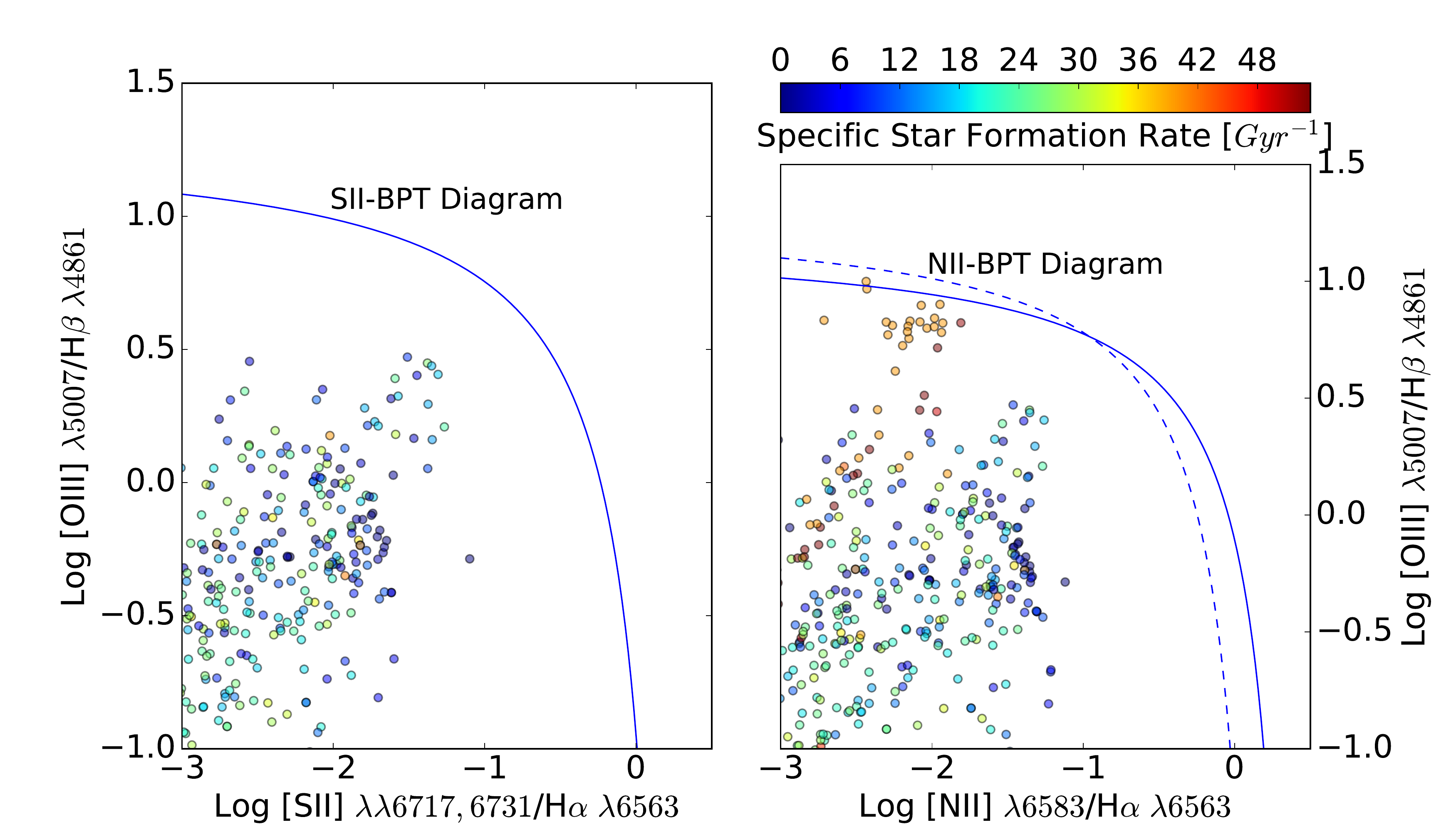}
\caption{\sii{} and \nii{} BPT diagrams of the emission line ratios coloured by specific star formation rate. The AGN line in the work by \citet{2003MNRAS.346.1055K} is shown as a solid line and the work by \citet{2001ApJ...556..121K} is shown as a dashed line with AGN occupying the region above and to the right of the lines.}
\label{fig:bpt0}
\end{center}

\end{figure*}

\subsubsection{Composite Spectra}

\begin{table}
\begin{center}
\caption{Linear regression analysis between flux ($z=15$) and stellar mass}
  \begin{tabular*}{0.99\columnwidth}{@{\extracolsep{\fill}} ccccc}
    Filter & $\beta_1$ & $SE_{\beta,1}$ & $\beta_0$ & $\rm{R^2}$ \\
\hline
f125w & 3.0326 & 0.1595 & -43.6215  & 0.1802 \\
f140w & 1.4737 & 0.0270 & -28.6025  & 0.6443 \\
f160w & 1.4154 & 0.0245 & -28.0498  & 0.6705 \\
f115w & 1.2554 & 0.0299 & -31.2656  & 0.5169 \\
f150w & 1.4243 & 0.0249 & -28.1705  & 0.6657 \\
f200w & 1.2606 & 0.0168 & -26.9581  & 0.7730 \\
f277w & 1.2165 & 0.0145 & -26.8831  & 0.8110 \\
Bolometric & 1.1984 & 0.0137 & 34.8578  & 0.8233 \\   

    \hline
  \end{tabular*}
  \parbox[t]{0.99\columnwidth}{\textit{Notes:} Correlations are given between log flux (in units of $\rm{erg/s/cm^2}$) and log stellar mass for Hubble (top) and JWST (middle) IR filters at $z=15$ as well as the bolometric luminosity (in units of $\rm{erg/s}$). The columns show filter, slope, the standard error of the slope, zero-point, and R$^2$, respectively.  The functional form is assumed to be $\rm{log}\ f = \beta_{\rm{0}} + \beta_{\rm{1}} \rm{log\ M_\star}$. 
  }
\label{tab:slopeshubble}
\end{center}
\end{table}

\begin{table}
\begin{center}
\caption{Linear regression analysis between apparent magnitude ($z=15$) and stellar mass}
\begin{tabular*}{0.99\columnwidth}{@{\extracolsep{\fill}} ccccc}
  Filter & $\beta_1$ & $SE_{\beta,1}$ & $\beta_0$ & $\rm{R^2}$ \\
  \hline
J & -7.3825 & 0.4905 & 90.8858  & 0.1211 \\
H & -3.4369 & 0.0563 & 49.5863  & 0.6938 \\
K & -3.0903 & 0.0389 & 46.7610  & 0.7931 \\

  \hline
\end{tabular*}
\parbox[t]{0.99\columnwidth}{\textit{Notes:} Correlations are given for various filters, redshifting the spectra to $z=15$.  The columns show filter, slope, the standard error of the slope, zero-point, and R$^2$, respectively.  The functional form is assumed to be $m_x = \beta_0 + \beta_1 \log M_\star$.
}

\label{tab:slopes}
\end{center}
\end{table}

Table \ref{tab:slopeshubble} shows the slope of the magnitudes in various HST and JWST filters as a function of stellar mass for objects observed at $z=15$. HST filter f125w and JWST filter f115w examine wavelengths that lie partially beyond the Lyman break in the rest frame and exhibit low correlation. For the other filters, linear regression analysis demonstrates $\rm{R^2}$ varies between 0.64 and 0.77 in each filter, implying relatively large variation of individual halo luminosity when summed over the frequency space corresponding to an individual spectra. The locus of points used to generate the regression of the bolometric luminosity are presented in the bottom panel of Fig. \ref{fig:lum1}. Fig. \ref{fig:lum2} shows corresponding countor histograms for HST filter f160w and JWST filter f277w which have the highest infrared mean wavelengths and are thus the most useful filters in either telescope for studying linear trends in this epoch. While linear fits seem to be appropriate for haloes with lower stellar masses, we see a tendency for the fits to overestimate the mean flux and luminosity of our spectra for ${\rm M}_\star > 10^6\ {\rm M}_\odot$. When narrower bands corresponding to the Vega colour designations are used, the $\rm{R^2}$ values are generally higher as shown in Table \ref{tab:slopes} for measures of absolute magnitude at $z=15$. This can be somewhat explained by the variation in emission lines demonstrated in part by Fig. \ref{fig:lines0} and the tendency for larger bands to include more lines and thus more variability. Additionally, J, H and K filters demonstrate a higher tendency towards a linear relationship with higher wavelength. The slope of absolute magnitude with respect to solar mass generally becomes more gradual (less negative) with wavelength.

Fig. \ref{fig:spectra0} shows composite spectra for four ranges of stellar mass as well as gas mass fractions below and above half the mean baryon fraction ($\Omega_{\rm b}/\Omega_{\rm M}$). Each plot shows the mean and standard deviation of the the spectra among the stated sample with each composite spectra normalized to 1500~\AA(rest), where the overall normalization can be inferred from the luminosities in Fig. \ref{fig:lum1}. Like the distribution of emission line strengths, the spread of the spectra appears to be inversely related to the mass of the halo. This may be partially explained by the sampling bias due to small numbers of star particles in small haloes. For larger haloes ($M_\star = 10^6  - 10^7\ \rm{M_\odot}$) with several hundred or more stellar particles, the wider standard deviation may be related to greater variability in the metallicity and temperature of the CSM about each halo and less variability in the centre of the clusters containing the largest haloes.

Fig. \ref{fig:bpt0} shows the Baldwin, Phillips and Terlevich \citep[BPT;][]{1981PASP...93....5B} diagrams of both (\oiii{} $\lambda$5007)/H$\beta$ to (\sii{} $\lambda\lambda$6717,6731)/H$\alpha$ and (\oiii{} $\lambda$5007)/H$\beta$ to (\nii{} $\lambda$6583)/H$\alpha$. BPT diagrams are customarily used to delineate the boundary between normal star-forming galaxies and active galactic nuclei. In our data, all points lie below the AGN boundary for \sii{} \citep{2003MNRAS.346.1055K} and the AGN boundary for \nii{} \citep{2001ApJ...556..121K}. Though some objects graze the boundary in the \nii{} diagram, there is some evidence that this is a feature of star-formation dominated spectra for high-redshift objects \citep{2013ApJ...774L..10K}.

\begin{figure}
\begin{center}
\includegraphics[scale=.36]{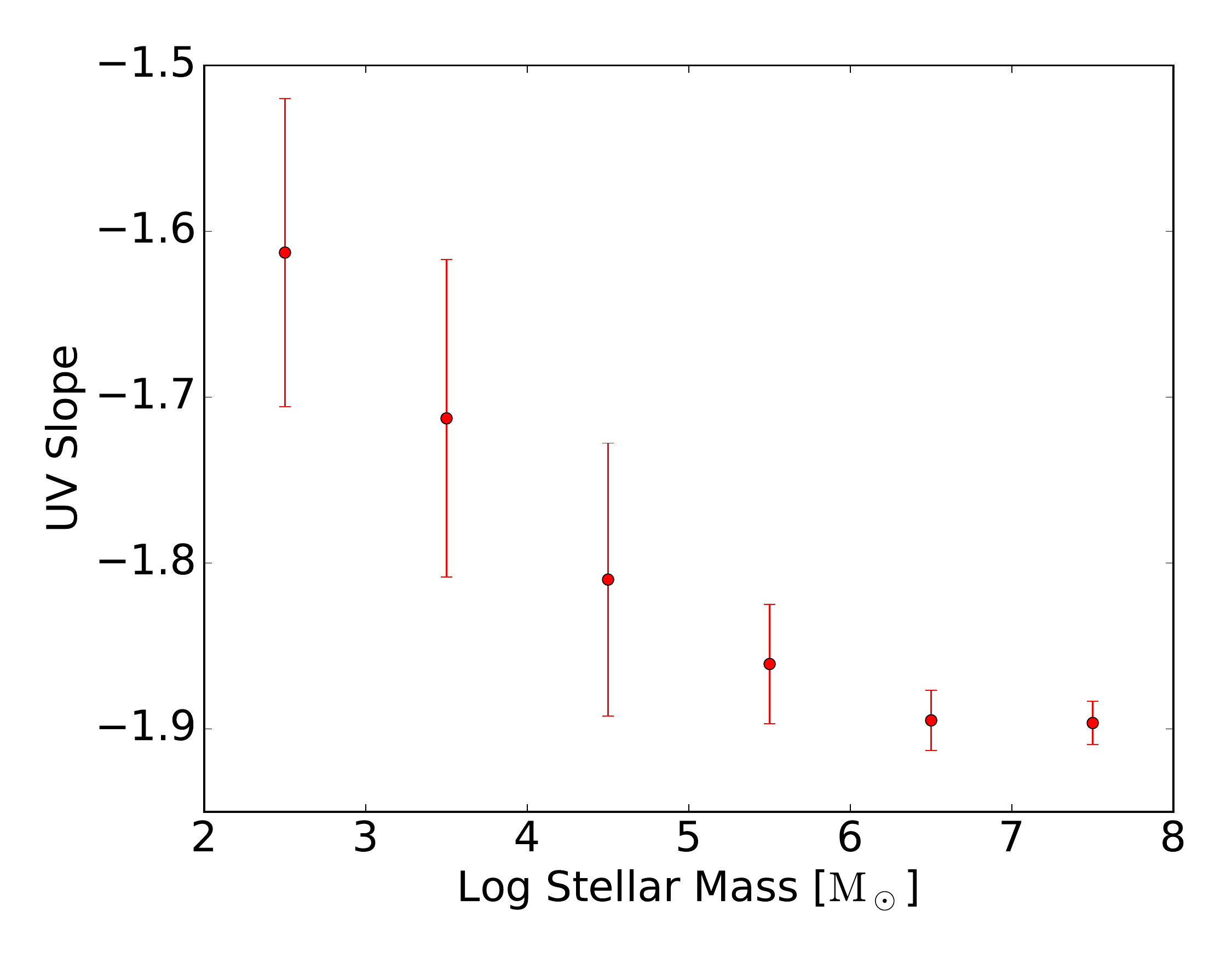}
\caption{The UV slope of averaged spectra versus stellar mass, showing the galaxies becoming bluer with increasing stellar mass up to $10^6 \Ms$.  This steepening of the UV slope occurs because all galaxies above this mass range host active star formation, whereas only a fraction of less massive galaxies host active star formation.}
\label{fig:UVslope}
\end{center}

\end{figure}
\begin{figure*}
\begin{center}
\includegraphics[scale=.37]{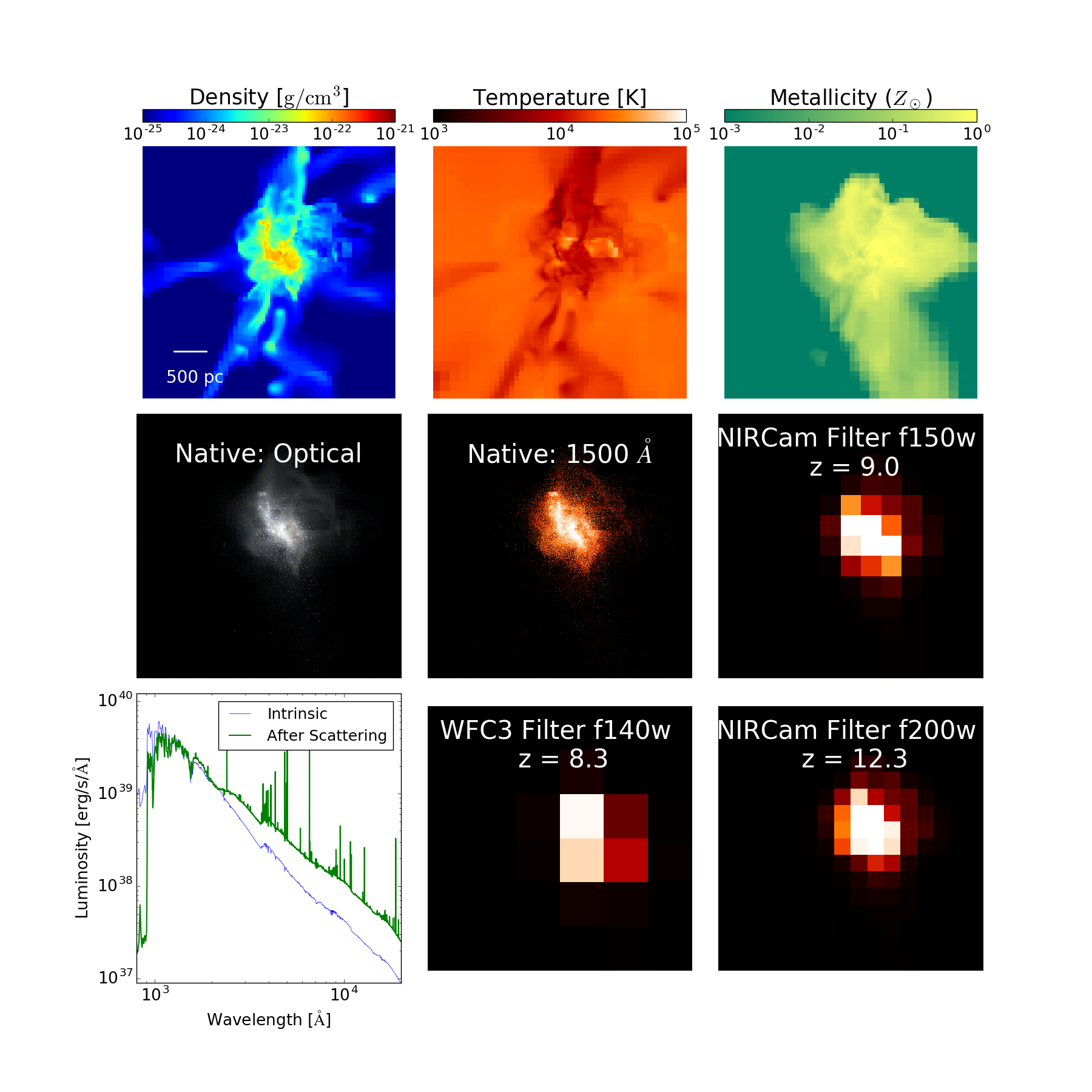}
\caption{Synthetic imaging and spectrum of Halo A $(M_{\rm{tot}} = 1.05 \times 10^9\ \rm{M_\odot}$, $M_\star = 2.04 \times 10^7\ \rm{M_\odot}$). Top Row: Density-weighted projections of density, temperature, and metallicity.  Middle Row: composite of optical frequency images, a monochromatic image of the halo at 1500 \AA\, and the same image as seen through the JWST NIRCam F150W filter (1.5~\micron). Bottom Row: intrinsic stellar spectrum (thin blue line) and processed galactic spectrum with dust and gas absorption, re-emission, and emission lines (thick green line), the 1500 \AA\ image as seen through Hubble's WFC3 F140 filter (1.4~\micron) and JWST NIRCam F200W filter (2.0~\micron) with the latter having a lensing magnification factor $\mu = 10$.}
\label{fig:HaloA}
\end{center}
\end{figure*}

\begin{figure*}
\begin{center}
\includegraphics[scale=.30]{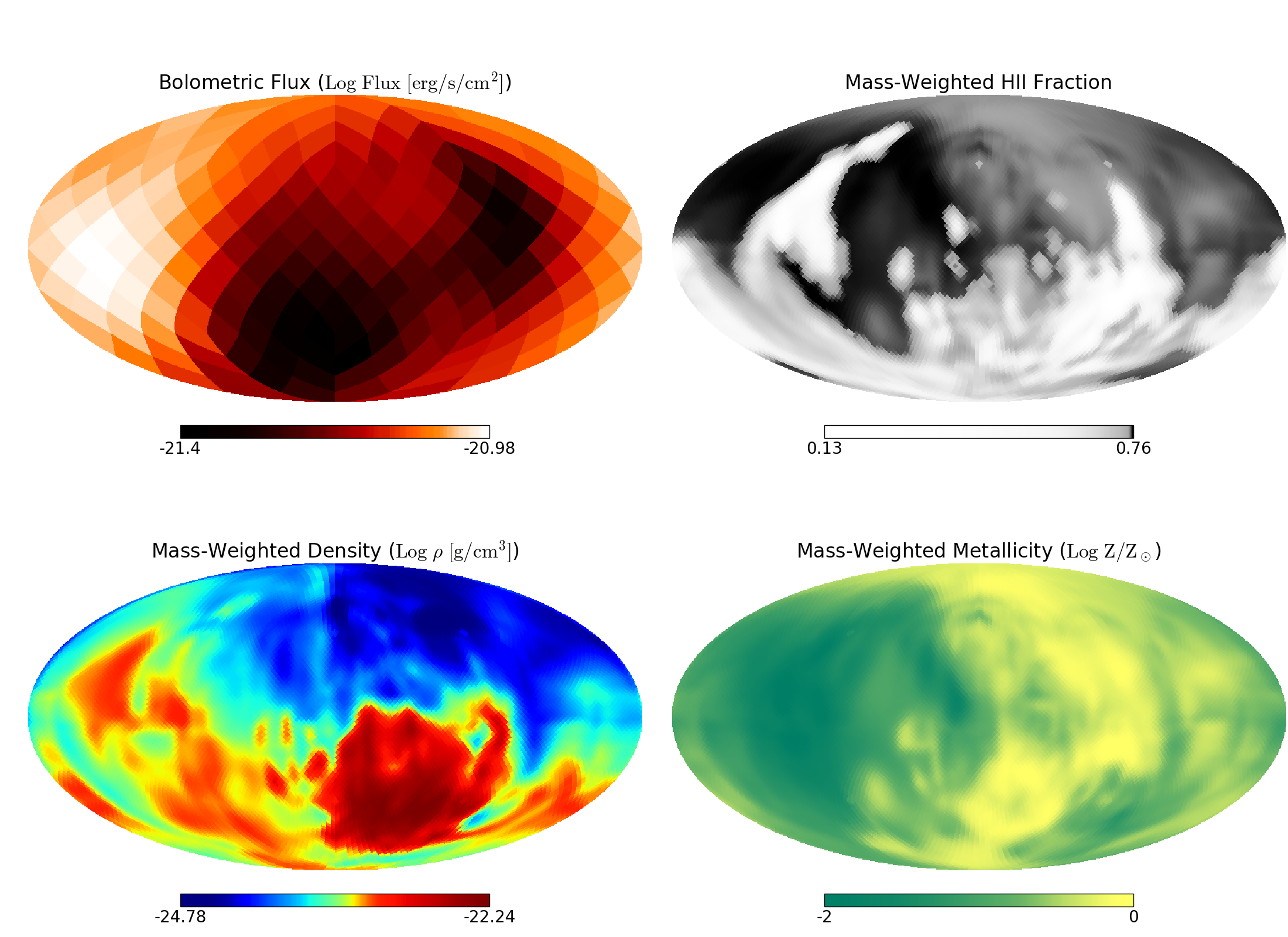}
\caption{Variations of emergent flux and gas properties of Halo A along different lines of sight.  Top row: Ecliptic view of total flux for Halo A at $z = 15$ (left) and the mass-averaged \hii{} fraction (right). Bottom row: Mass-averaged density and metallicity along the same normals integrated to the virial radius of Halo A.}
\label{fig:mproj192}
\end{center}
\end{figure*}

\begin{figure*}

\begin{center}
\includegraphics[scale=.37]{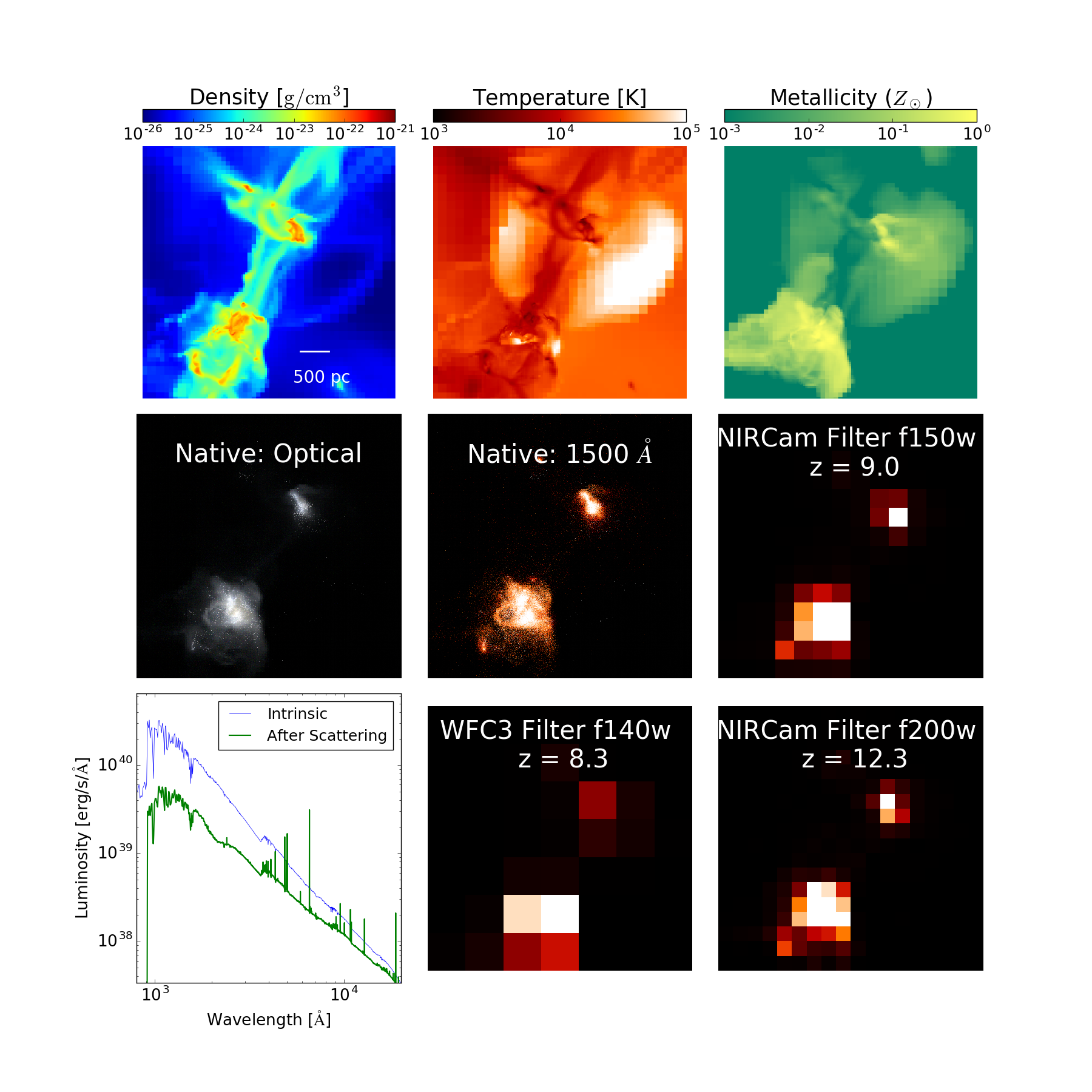}
\caption{Halo B $(M_{\rm{tot}} = 1.62 \times 10^9\ \rm{M_\odot}$, $M_\star = 3.4 \times 10^7\ \rm{M_\odot}$) plotted in the same manner as Halo A in Fig. \ref{fig:HaloA}.}
\label{fig:HaloB}
\end{center}
\end{figure*}

The average spectral slope in the range 1200--2500\ \AA{}(rest) is plotted in Fig. \ref{fig:UVslope} assuming flux follows a power law relationship with respect to wavelength ($f_\lambda \propto \lambda^\beta$). UV slopes are calculated using a linear regression on the full SEDs for the indicated mass bins between 1200 and 2500 \AA (rest). As discussed, bins of low stellar mass can include populations of haloes with relatively old and red stellar populations whereas bins of higher stellar mass include a large distribution of young and older stellar populations, resulting in a fairly consistent mean stellar age for our entire sample of large galaxies. Thus, haloes with stellar masses between $10^7$ and $10^8\ \rm{M_\odot}$ ($n = 5$) have relatively low variability and average UV slopes of around $\beta =  -1.85$. The intrinsic {\sc FSPS} galactic spectra for the same sample is between $\beta = -1.85$ and --1.90 implying that our method resulted in a slightly shallower slope for our largest objects. Other mass bins show little to no change in the UV slope after applying our pipeline.

\subsection{Individual haloes}
\label{IndiHalo}

We present two cases from the Renaissance Simulations to demonstrate the impact and utility of our dust scattering and emission line method. Relevant statistics for the haloes discussed are provided in Table \ref{tab:Halolist}. 

\begin{table}
  \centering
  \caption{Individual halo properties}
  \begin{tabular*}{0.99\columnwidth}{@{\extracolsep{\fill}} cccccc}
    Halo & $\log M_{\rm tot}$ & $\log M_\star$ & $f_{\rm gas}$ & $Z_\star$ & $\log L_{\rm tot}$\\
    & [\Ms] & [\Ms] & & [\zsun] & [erg s$^{-1}$]\\
    \hline
    A & $9.02$ & $7.31$ & 0.164 & 0.410 & $42.88$ \\
    B & $9.21$ & $7.57$ & 0.133 & 0.440 & $43.03$ \\
    \hline
  \end{tabular*}
  \parbox[t]{0.99\columnwidth}{\textit{Notes:}
    The columns show halo mass, stellar mass, gas mass fraction, mass-averaged stellar metallicity, and total bolometric luminosity.
  }
\label{tab:Halolist}
\end{table}

Halo A (Fig. \ref{fig:HaloA}) is the third most massive halo in our simulation with a total mass of $1.05 \times 10^9 \ \rm{M_\odot}$ and a stellar mass of $2.04 \times 10^7\ \rm{M_\odot}$. We see from the density projection (top left panel) that the halo is centred on a single large galaxy surrounded by several satellite galaxies and sub-haloes. Halo A  has a mass-weighted mean metallicity of 0.242 $\rm{Z_\odot}$, which corresponds to the presence of enough dust to attenuate the galactic spectra. Defining the circum-stellar medium (CSM) as the medium within the virial radius of a halo that exists about but not necessarily between the halo's stellar population, we observe that the metallicity distribution (top right) is uneven with large volumes of the CSM un-enriched.  The optical composite (middle left) shows stellar populations mostly concentrated in the main galaxy within regions of high gas density. We also see a smaller diffuse population of star particles towards the bottom of the image that have been stripped from the main body of the halo into a region of relatively low gas density. In the SED (bottom left), higher gas densities contribute to a significant degree of frequency-dependent scattering and reprocessing of the intrinsic stellar spectra towards lower frequencies. Due to diffuse emission \citep[cf.][]{2009A&A...502..423S}, the bolometric luminosity of the galaxy after processing through our pipeline exceeds the intrinsic luminosity due solely to stellar sources by 0.64 mag (80\% brighter). We also observe that Halo A would be discernible above the noise at $z = 8.3$ using WFC3 and would exhibit some structure through NIRCam at $z = 12$ assuming gravitational lensing by a factor of $\mu = 10$.

{\sc Hyperion} allows for the calculation of the flux incident onto a plane of arbitrary orientation. By using {\sc HEALPix} to generate equally spaced normals about a sphere, we generate a composite image of flux from a single galaxy as a function of the observer's viewing angle. Line integration of physical quantities along those normals from the center of the halo to its virial radius produces corresponding plots of density and metallicity for comparison.

Fig. \ref{fig:mproj192} shows the integrated flux, \hii{} fraction, mean density, and metallicity as a function of viewing angle about Halo A normalized at $z = 10$ without IGM absorption. Depending on the viewing angle, the total integrated flux at the observer from Halo A varies by a factor of $\sim$ 3 emphasizing the anisotropic nature of modelling ISM and CSM attenuation and scattering despite a relatively central arrangement of stars and gas. Furthermore, line integrals of mass-weighted mean density through the center of the halo vary by more than two orders of magnitude depending on the normal direction. Normals along high mean density correspond to lower bolometric flux, but normals along lower mean densities do not result in the highest flux at the observer. Since this halo exhibits heavy scattering and a 0.64 magnitude increase in bolometric luminosity due to diffuse emission, the brightest flux corresponds to regions of intermediate density where enough gas is present to contribute to emission, but not enough to exhibit self-shielding.
Mass-weighted metallicity shows a slightly more complicated relationship as some regions of both high and low densities are metal-enriched and the most metal poor directions have intermediate-gas densities. The brightest normals appear to be therefore correlated to the lowest gas metallicities implying a lack of prior star formation in those regions. The plot of \hii{} fraction confirms the presence of confined \hii{} regions that are observed in Fig. \ref{fig:HaloA} as the hottest pockets of gas in the temperature projection plot. These \hii{} regions appear to be powering diffuse emission and the greater flux. Taken together, confined \hii{} regions around young stars in intermediate-density metal-poor gas are producing diffuse emission that result in a significantly brighter flux at the observer than could be explained with intrinsic stellar spectra, but with great variability with respect to viewing angle. 

A system of merging galaxies within Halo B is shown in Fig. \ref{fig:HaloB}. This halo is notable for its high mass-weighted mean stellar metallicity of 0.440 $\rm{Z_\odot}$ and contains the largest total mass of any halo in our simulation ($1.62 \times 10^{9}\ \rm{M_\odot}$). Like Halo A, Halo B is plainly visible through HST and JWST assuming a gravitational lensing by a factor of $\mu = 10$. It appears as two distinct visible galaxies connected by a faint filament of stellar clusters. The larger galaxy in the lower portion of the projections is metal-enriched, showing near Solar metallicity in the brightest regions and metallicities above $10^{-1}\ \rm{Z_\odot}$ in its CSM. The smaller galaxy is also metal-enriched but the adjacent filamentary structures are notably metal poor despite the presence of stars. This implies that the stars and structure of the filament are relatively young. 

The hottest regions of the temperature projection ($T \geq 10^5$~K) trace multiple supernova remnants centred in a region of disrupted gas that includes the smaller galaxy. This further implies that the halo has been disrupted into its current configuration from a set of two more cohesive merging galaxies with established metal-enriched stellar populations. Temperature projections also indicate the presence of \hii{} regions in the larger galaxy, which also appear as diffuse radiation in the 1500~\AA(rest) image. The intrinsic bolometric luminosity of the stellar sources peak at approximately $3 \times 10^{40}\ \rm{erg\ s^{-1}}$ \AA$^{-1}$\ at 935~\AA (rest), but is attenuated by the gas and dust to a peak of around $6 \times 10^{39}\ \rm{erg\ s^{-1}}$ \AA$^{-1}$\ at 1050~\AA (rest) with roughly similar attenuation for all frequencies simulated.

\section{Discussion}
\label{discussion}

Due to the anisotropic nature of dust and gas scattering and absorption, a fully three dimensional model is required for representative modelling of the spectral energy distribution of a galaxy. Furthermore, the effect of viewing angle is shown to produce large variation in flux at the observer in even in relatively well-ordered systems. This may therefore result in large uncertainties when interpreting observations of the first galaxies, however most of the overall variability in the results is confined to haloes with total masses below $10^8\ \rm{M_{\odot}}$. Above this mass, results converge in colour-colour diagrams, equivalent and line ratio plots, and plots of composite spectra.

Some of this convergence may be explained by the similarly ionized cosmological environment about large haloes with relative high star formation rates and some of it may be explained by deeper gravity wells acting against the tendency for haloes to become gas-poor through photo-evaporation or gas blowout. Conversely, the divergence at low mass can be explained by bursts of star formation and diversity in the \hii{} regions around galaxies.

Our picture of radiation through galaxies in the EoR is incomplete however. Our method calculates the photoionization in the \hii{} regions around clusters by assuming the cluster radiates as a single source. Because the simulation forms clusters that have a minimum mass of $1000\ \rm{M_\odot}$, the nature of the ISM between stars within a cluster is not simulated directly by our model. Furthermore, we do not apply the Monte-Carlo process directly to the scattering and attenuation of emission lines so we do not capture the effect of an uneven medium at larger scales either.

The size of our stellar cluster particles implies the assumption that stars do not diffuse through a galaxy during the course of our simulation. This has the effect of both over-estimating the flux of photons around clusters and underestimating the flux throughout the rest of the ISM for older clusters. This may also have an impact on the star formation rate and the distribution of supernovae. We also do not include the spectra or luminosity of Population III stars or accreting black holes in the model at this time so we cannot speak to their contribution to the photometry and spectra of early galaxies.

Since our spectra are simulated for a closed volume about a halo, we do not capture the effect of the outer circum-galactic medium or the intergalactic medium (IGM). While we do account for cosmological effects, several intermediate and foreground effects may cause observations of galaxies simulated by this work to differ materially from observed galaxies that are otherwise similar in nature. We encourage the use of an appropriate model for the IGM in comparisons. For example, one could use the galaxy spectrum as the background of a ray-trace through the IGM, as used by {\sc Trident} \citep{2016arXiv161203935H}.  This will include all important effects on the spectrum outside of the virial radius, including absorption from \hi{} and metal species as well as galaxy foreground effects.

\subsection{Comparable Works}

\subsubsection{Zackrisson et al. (2013)}

Using stellar SEDs from the {\sc Yggdrasil} population synthesis code \citep{2011ApJ...740...13Z} and {\sc Cloudy}, \citet{2013ApJ...777...39Z} explore the evolution of the UV slope of galaxies over time and with respect to H$\beta$ EW for a range Lyman continuum escape fractions. They argue that for high escape fractions, galaxies with $M_\star \geq 10^7\ {\rm M}_\odot$ should be detectable with JWST up to $z \simeq 9$ where our entire sample of galaxies with $M_\star \geq 10^7\ {\rm M}_\odot$ is visible to redshift 12 with $\mu = 10$. They note that the contribution of dust is harder to determine and suggest the use of ALMA to constrain the IR dust emission peak of observations. 

\subsubsection{Cen and Kim (2014)}

\citet{2014ApJ...782...32C} use {\sc Enzo} to calculate a zoom-in region about a $3 \times 10^{14}\ {\rm M}_\odot$ halo. They produce stellar SEDs with {\sc Gissel} \citep{2003MNRAS.344.1000B} and use the code {\sc Sunrise} \citep{2006MNRAS.372....2J,2010MNRAS.403...17J} to model dust absorption and scattering. Emission lines are added to the spectra of young star clusters using the code {\sc Mappingsiii} \citep{2010MNRAS.403...17J} by assuming a constant star formation rate for 10 Myr. To motivate upcoming observations by ALMA, they explored peak IR wavelengths, luminosity functions, and FUV-NUV colors and found their results to be in good agreement with observations. This work concurs with their result showing markers of bursty star formation in low mass haloes. 

\subsubsection{Wilkins et al. (2016)}

Using the {\sc GADGET} Bluetides Simulation \citep{2016MNRAS.455.2778F}, \citet{2016MNRAS.460.3170W} applied {\sc Cloudy} to generate emission lines lines for EoR galaxies. They compared the effect of different stellar population synthesis models on the UV slope and found a $\sim 0.2$ variation in $\beta$. They predict somewhat steeper UV slopes (-2.4 to -2.5 for FSPS) in their sample using $\beta = 1.8 \times (m_{\rm{fuv}} - m_{\rm{nuv}}) - 2.0$ and FUV and NUV magnitudes whereas our study directly calculates UV slopes using a regression (Fig. \ref{fig:UVslope}). Their study of the effect on redshift also finds that galaxies became more blue at higher redshift corresponding to younger stellar populations. Our work concurs with redder colours in the higher mean stellar age bins, but is confined to a single redshift and therefore does not track the evolution of the distribution of colours as galaxies age.

\subsubsection{Cullen et al. 2017}

A recent submission by \citet{2017arXiv170107869C} explores the effect of dust on galaxies in the {\sc Gadget} First Billion Years Project \citep{2015MNRAS.451.2544P}. Stellar population synthesis is calculated using {\sc BPASSv2} \citep{2016MNRAS.462.3302E} and emission lines are calculated using {\sc Cloudy}. In contrast to this work and the work of \citet{2014ApJ...782...32C}, dust attenuation is accomplished by comparing a suite of different analytic models rather than a Monte Carlo process. One of their most consistent dust attenuation laws is reported as $A_{1600} = 2.10^{1.9}_{-0.3}(\beta + 2.52)$ at $z = 5$ where $\beta$ is the UV slope. They note that this implies a higher star formation rate than previously deduced and appeal to ALMA for confirmation.

\subsection{Applications}

Our data set includes synthetic observations that would be visible with current telescopes, but the vast majority of our mock observations are too far and too dim for the current generation of hardware. This work therefore extends our measures and figures to regimes that await validation by JWST and other future telescopes. Furthermore, our sample is large enough for some measures to be compared to a statistically significant body of observed EoR galaxies and therefore serves as a prediction of trends and distributions of colour-colour plots, luminosity, and emission lines. We also see utility in the use of our investigation as a means for the preliminary categorization and characterization of future observations.

\subsection{Future Enhancements}

The JWST team is developing an image processing calculator\footnote{https://demo-jwst.etc.stsci.edu/} that processes raw source photometric data into synthetic JWST results. The potential exists to integrate the source code into this pipeline to refine the analysis and ensure better confluence with observational data beyond the methods explored thus far. 

Emission line calculations are currently limited to an augmented luminosity calculation that needs to be further refined and calibrated. A method that directly simulated the scattering of the emission lines throughout the ISM is preferred. 

The current method neglects the inclusion of Population III stellar spectra in its first iteration due to the exclusion of this type of star from the isochrones available in {\sc FSPS}. Theoretical source spectra are available in {\sc Yggdrasil} and will be integrated into the second iteration. Additionally, by including composite spectra models of accretion disks and the region around black holes, the potential exists to extend this method to simulations of galactic nuclei and starbursts about black holes.

Additionally, with the development of synthetic photometry and spectra pipelines, tracking the evolution of the flux and spectra over time presents a novel line of investigation and a natural extension of the work completed to this point. With few extensions, all the results presented in the previous section may be converted into a time-series analysis by connecting data from each simulation data output.

\section{Conclusions}   

With deep field observations using the Hubble Frontier Fields and the forthcoming JWST enabling the collection of a statistically significant sample of galaxies at high redshift, cosmological simulations of the Epoch of Reionization offer an opportunity to make predictions. We employ results from the ``rare peak'' zoom-in region of the Renaissance Simulations to stage a Monte Carlo photon simulation of continuum radiation through dust and gas as well as calculations of nebular emission lines. We generate synthetic photometry and spectra for two of our largest individual galaxies and provide photometric measures of the entire aggregate sample. Our study of the larger individual galaxies reveals the following insights:
\begin{enumerate}
\item Dust and gas attenuation is non-isotropic and wavelength-dependent for some haloes. Our first example shows that flux versus inclination is not necessarily a function of column density for irregular galaxies. The distribution of \hii{} regions and the age of bursts of star formation may result in viewing angles where a galaxy may appear several times brighter.
\item The intrinsic spectra produced by a stellar population synthesis model may understate the final luminosity of a galaxy after the calculation of gas and dust scattering in addition to the effect of \hii{} regions in some situations. 
\end{enumerate}

Our study of the entire sample of star-containing haloes reveals the following trends:
\begin{enumerate}
\item haloes with the lowest total mass have the highest variability in bolometric luminosity (4 orders of magnitude). Likewise, haloes with the lowest stellar mass have the greatest variability in \oiii{}/H$\beta$, \oiii{} EW and \ciii{} EW. This is due to cycles of star formation feedback having a significant impact on the gas in the smallest haloes. Ly$\alpha$ equivalent widths are less than 1\AA\ and exhibit an inverse relationship with $M_\star$ and electron fraction.
\item The slope of luminosity versus stellar mass becomes shallower and more linear with the mean wavelength of a filter band. Variations in the spectra between haloes are least between the highest stellar mass haloes and greatest between our lowest stellar mass haloes.
\item Our method causes a small decrease in $\rm{J_{200w}-J_{277w}}$ colour and a up to a $\sim 0.5 $ decrease in $\rm{J_{150w}-J_{277w}}$ colour versus the intrinsic FSPS stellar spectra for larger-mass . 
\item UV slopes decrease as stellar mass increases for halo with stellar masses between $10^{3}$ to $10^8\ \rm{M_\odot}$. Our method results in shallower UV slopes than the intrinsic stellar population synthesis spectra for the highest mass objects.
\end{enumerate}

We have shown the impact of anisotropic, non-homogeneous dust and gas distributions on mock photometry and predictions for galactic spectra. Our treatment illustrates a method for the characterization of future observations of the early Universe and provides a large sample of mock observations that demonstrate physically-motivated trends in emission lines, colours, and luminosity from a large, representative cosmological simulation.

\section*{Acknowledgements}
KSSB acknowledges support from the Southern Regional Education Board doctoral fellowship.
JHW acknowledges support from National Science Foundation (NSF) grants
AST-1333360 and AST-1614333 and Hubble theory grants HST-AR-13895 and
HST-AR-14326.  MLN was supported by NSF grant AST-1109243.  BWO was
supported in part by NSF grants PHY-1430152 and AST-1514700, by NASA grants NNX12AC98G, NNX15AP39G, and by
Hubble Theory Grants HST-AR-13261.01-A and HST-AR-14315.001-A.  The
simulation was performed using \textsc{Enzo} on the Blue Waters
operated by the National Center for Supercomputing Applications (NCSA)
with PRAC allocation support by the NSF (award number
ACI-0832662). This research is part of the Blue Waters
sustained-petascale computing project, which is supported by the NSF
(award number ACI 1238993 and ACI-1514580) and the state of Illinois. Blue Waters is a
joint effort of the University of Illinois at Urbana-Champaign and its
NCSA.  This research has made use of NASA's Astrophysics Data System
Bibliographic Services.  Analysis was performed on XSEDE's Maverick
resource with XSEDE allocation AST-120046.  The majority of the
analysis and plots were done with \textsc{yt} and \textsc{matplotlib}.
\textsc{Enzo} and \textsc{yt} are developed by a large number of
independent researchers from numerous institutions around the
world. Their commitment to open science has helped make this work
possible.

\bsp
\label{lastpage}
\end{document}